\documentclass[draftcls, onecolumn]{IEEEtran}
\ifCLASSINFOpdf
\else
\fi
\usepackage{slashbox}
\usepackage{graphicx}
\usepackage{amsmath}
\usepackage{amssymb}
\usepackage{graphics}
\usepackage{latexsym}
\usepackage[dvips]{epsfig}
\usepackage[nospace]{cite}
\usepackage{subfigure}
\usepackage{setspace}
\usepackage{tabularx}
\usepackage{color}
\usepackage{arydshln}

\newtheorem{Definition}{Definition}
\newtheorem{Lemma}{Lemma}

\newtheorem{Proposition}[Lemma]{Proposition}
\newtheorem{Theorem}{Theorem}

\newtheorem{Example}{Example}
\newtheorem{Remark}{Remark}

\newcommand\blue[1]{\textcolor{blue}{#1}}
\newcommand\magenta[1]{\textcolor{magenta}{#1}}

\allowdisplaybreaks[1]

\def\rank{{\rm {rank}}}
\def\diag{{\rm {diag} }}

\begin{document}
%
\title{Optimal Streaming Codes for Channels with Burst and Arbitrary Erasures}


\author{Silas L.~Fong, Ashish Khisti, Baochun Li, Wai-Tian Tan, Xiaoqing Zhu, and John Apostolopoulos%
\thanks{This paper was presented in part at 2018 IEEE International Symposium on Information Theory.}
\thanks{S.~L.~Fong, A.~Khisti and B.~Li are with the Department of Electrical and Computer Engineering, University of Toronto, Toronto, ON M5S 3G4, Canada  (e-mails: \texttt{silas.fong@utoronto.ca}, \texttt{akhisti@ece.utoronto.ca}, \texttt{bli@ece.utoronto.edu}).}
\thanks{W.-T.~Tan, X.~Zhu and J.~Apostolopoulos are with Cisco Systems, San Jos\'e, CA 95134, USA.}
}


%


\maketitle

\begin{abstract}
This paper considers transmitting a sequence of messages (a streaming source) over a packet erasure channel. In each time slot, the source constructs a packet based on the current and the previous messages and transmits the packet, which may be erased when the packet travels from the source to the destination. Every source message must be recovered perfectly at the destination subject to a fixed decoding delay.
We assume that the channel loss model introduces either one burst erasure or multiple arbitrary erasures in any fixed-sized sliding window. Under this channel loss assumption, we fully characterize the maximum achievable rate by constructing streaming codes that achieve the optimal rate.
In addition, our construction of optimal streaming codes implies the full characterization of the maximum achievable rate for convolutional codes with any given column distance, column span and decoding delay. Numerical results demonstrate that the optimal streaming codes outperform existing streaming codes of comparable complexity over some instances of the Gilbert-Elliott channel and the Fritchman channel.
\end{abstract}


%
\IEEEpeerreviewmaketitle

\section{Introduction} \label{Introduction}
Low-latency video conferencing has been a cornerstone for communication and collaboration for individuals and enterprises. The advent of 5G networks promises to make high-throughput at low-latency ubiquitous. This enables new applications such as high-quality video conferencing, virtual reality (VR) and Internet-of-things (IoT) applications including vehicle-to-vehicle communication and mission-critical machine-type communication~\cite{5GPPP}. At the core of these important applications is the need to reliably deliver packets with low latency. Packet losses at the physical layer and the network layer are inevitable, which may be caused by unreliable wireless links or congestion at network bottlenecks.
In order to alleviate the effect of packet losses on applications that are run over the Internet, two main error control schemes have been implemented at the data link and transport layers: Automatic repeat request (ARQ) and forward error correction (FEC).

For long-distance low-latency communication, it is not suitable to use ARQ schemes for error control because each retransmission incurs an extra round-trip delay. More specifically, correcting an erasure using ARQ results in a 3-way delay (forward + backward + forward), and this aggregate (3-way) delay including transmission, propagation and processing delays is required to be lower than~150 ms for interactive applications such as voice and video according to the International Telecommunication Union~\cite{onewayTransTime} \textcolor{black}{(see~\cite{StockhammerHannuksela2005} for an overview of the ubiquitous H.264/AVC video coding standard)}. This aggregate delay makes ARQ impractical for communication between two distant points with aggregate delay larger than 150~ms. For example, ARQ cannot be used for communication between two diametrically opposite points on the earth's circumference because the corresponding propagation delay alone is at least 200~ms~\cite{BKTAmagazine17}.

For short-distance low-latency communications in the Tactile Internet, the next evolution of IoT, whose round-trip latency is required to be less than 1 ms~\cite{5GPPP}, using ARQ schemes at the transport layer for error control is an inefficient use of precious time resources because the time budget allocated for retransmissions could instead be used for processing data at end users or data processing servers. Consider the example of remotely controlling a critical device where a sensor wants to communicate with an actuator in real time through a control server with round-trip latency less than 1 ms as illustrated in~\cite[Fig.~3]{5GPPP}. The latency goals for processing delay at the terminals, transmission delay over the air interfaces between the terminals and the control server and data processing delay at the control server are 0.3 ms, 0.2 ms and 0.5 ms respectively. If an ARQ scheme is used for error control, then retransmissions compete the precious time resources with data computation at the terminals and the control server.

On the contrary, forward error correction (FEC) schemes are amenable to low-latency communications because no retransmission is required. Instead of using retransmissions to achieve high reliability, FEC schemes increase the correlation among the transmitted symbols by adding redundant information. In other words, FEC schemes avoid the extra round-trip latency needed by retransmissions at the expense of the extra processing time spent on adding and removing redundant information at end users. Since FEC schemes inject redundancy at a constant rate while retransmissions inject redundancy at a highly non-uniform rate, FEC rather than ARQ schemes are more suitable for controlling delay for low-latency communications.

In order to search for FEC codes at the transport layer which are suitable for low-latency communications over the Internet, we are motivated to investigate the fundamental limits of low-latency streaming codes with FEC.

\subsection{Motivation of Studying Packet-Erasure Channel}\label{subsecMotivation}
In practice, packet losses experienced at the network layer can be well approximated by statistical models~\cite{HasslingerHohlfeld2008,HGH2008}, including the well-known Gilbert-Elliott (GE) channel~\cite{Gilbert1960,Elliott1963} and its generalization the Fritchman channel~\cite{Fritchman1967}. In order to find good FEC codes for error correction at the transport layer, it would be ideal if we could find the maximum achievable rate of a statistical model under a low decoding latency constraint and a given target error rate. However, characterizing such a rate over a statistical channel seems intractable. Therefore, we are motivated to study other simplified channel models that provide useful approximations to practical low-latency communications over the Internet.

In this paper, we focus on a packet-erasure channel model that introduces both burst and arbitrary errors. In any window of a fixed size, we assume that the channel introduces either a burst erasure or multiple arbitrary erasures. Although this channel model is not statistical, it has been shown in~\cite{BPKTA17} that streaming FEC codes that correct both burst and arbitrary erasures can significantly outperform traditional streaming FEC codes that correct only one type of erasures (either burst or arbitrary) for both the GE channel and the Fritchman channel.

\subsection{System Model}\label{subsecChannelModel}
In order to describe the existing results for the packet-erasure channel model, we would like to briefly describe the channel model. A formal description will appear later in the paper. The channel consists of a source and a destination. In each time slot, the source chooses a collection of~$k$ symbols destined for the destination and encodes the $k$ symbols into a collection of~$n$ symbols followed by transmitting the~$n$ symbols through the channel. The collection of~$n$ symbols transmitted in a time slot are either received perfectly by the destination or erased (lost). The fraction $k/n$ specifies the coding rate. We call the~$k$ symbols chosen by the source, the~$n$ symbols transmitted by the source and the~$n$ symbols received by the destination the \emph{source packet}, the \emph{transmitted packet} and the \emph{received packet} respectively. Since every low-latency application is subject to a tight delay constraint, we assume that every source packet generated in a time slot must be decoded with delay~$T$, i.e., within the future~$T$ time slots.

In order to capture the packet loss behavior over the Internet, we first consider the simple scenario where either one burst erasure with length no longer than~$B$ occurs or multiple arbitrary erasures with total count no larger than~$N$ occur on the discrete timeline. Since a channel that introduces any~$N$ arbitrary erasures can introduce any burst erasure of length~$N$, we assume without loss of generality (wlog) that
\begin{align}
B\ge N. \label{BgreaterThanN}
\end{align}
\textcolor{black}{In order to avoid triviality, we assume wlog that \begin{align}
B>0, \label{BgreaterThan0}
\end{align}
 or otherwise $B=N=0$ by~\eqref{BgreaterThanN} in which case no coding is needed to achieve the maximum coding rate~$1$ .
Similarly, a channel that introduces any burst erasure of a positive length can introduce one arbitrary erasure, hence we assume wlog that
\begin{align}
N\ge \begin{cases}
1 & \text{if $B>0$,}\\
0 & \text{otherwise.}
\end{cases} \label{NgreaterThan1}
\end{align}
}
In addition, we assume wlog that
\begin{align}
T\ge B, \label{assumptionWLOG*}
\end{align}
or otherwise a burst erasure of length~$B$ starting from a certain time slot would wipe out the source packet transmitted in the same time slot.
Under the erasure channel model described above, we are interested in characterizing the maximum coding rate $k/n$ for sending information over the channel such that every source packet can be perfectly recovered by the destination with delay~$T$. \textcolor{black}{In the rest of the paper, we assume wlog the following holds due to~\eqref{BgreaterThanN}, \eqref{BgreaterThan0}, \eqref{NgreaterThan1} and~\eqref{assumptionWLOG*}:
\begin{align}
T\ge B\ge N\ge 1. \label{assumptionWLOG}
\end{align}
}

\subsection{Related Work} \label{subsecRelatedWork}
Correcting burst erasures using convolutional codes has a long history starting in the late 1950's, and the achievable rates for convolutional codes that correct burst erasures have been discussed in numerous works including~\cite{Hagelbarger1959, WynerAsh1963, Massey1965, Forney1971}, but the optimality of the convolutional codes under delay constraints was not discussed until the work by Martinian and Sundberg~\cite{MartinianSundberg2004} in 2004. In~\cite{MartinianSundberg2004}, streaming codes for the special case $N=1$ are considered and the maximum achievable rate for convolutional codes over a channel that introduces only a single burst erasure (because $N=1$) was proved to be $\frac{T}{T+B}$. Various generalizations of the burst erasure model and the low-latency convolutional codes in~\cite{MartinianSundberg2004} have been proposed in~\cite{LeungHo2012, LQH2013, BKTA2013, AdlerCassuto2017}.

For a channel that introduces both burst and arbitrary erasures as described in Section~\ref{subsecChannelModel}, optimal convolutional codes with rate 1/2 were discovered in~\cite{BKTA2013} in 2013. Recently, it was proved by Badr et al.~\cite[Th.~1 and Th.~2]{BPKTA17} that the maximum achievable rate is bounded between $\frac{T-N}{T+B-N}$ and $\frac{T-N+1}{T+B-N+1}$ for any $(T,B,N)$.

\subsection{Main Contribution} \label{subsecContribution}
This paper studies the \emph{sliding window model} suggested in~\cite{BPKTA17} which generalizes the simple system model described in Section~\ref{subsecChannelModel}. Under this model,  we assume that either one burst erasure with length no longer than~$B$ occurs or multiple arbitrary erasures with total count no larger than~$N$ occur in any sliding window of size~$W$. If we set $W=\infty$, then the sliding window model reduces to the simple system model described in Section~\ref{subsecChannelModel}. Throughout this paper, we assume
\begin{align}
W\ge T+1 \label{WgreaterThanT}
\end{align}
 unless specified otherwise.
The assumption of the window size $W\ge T+1$ can be explained intuitively as follows ---
A source packet generated in a time slot must be decoded by the destination in~$T$ time slots, implying that the ``lifespan'' of each source packet is $T+1$. Setting the window size no smaller than the lifespan of a source packet enables us to investigate how the erasures within the lifespan of a source packet affects the recovery of the packet. Nevertheless, the case where $W<T+1$ will also be discussed in the sequel.

\textcolor{black}{Under the sliding window model, Badr et al.~\cite[Th.~1 and Th.~2]{BPKTA17} showed that the maximum achievable rate lies between
$\frac{T-N}{T+B-N}$ and $\frac{T-N+1}{T+B-N+1}$ for any~$(W, T, B, N)$, which is not a satisfactory result because the lower and upper bounds do not coincide for any~$(W, T, B, N)$.
The main result of this paper shows that the upper bound is indeed achievable, i.e., the maximum achievable rate equals
$\frac{T-N+1}{T+B-N+1}$ for any $(W, T, B, N)$.}
This generalizes the results in~\cite{MartinianSundberg2004,BKTA2013} and strengthens the result in~\cite{BPKTA17} (cf.\ Section~\ref{subsecRelatedWork}).
The exact statement of our main result will be stated in Section~\ref{subsecMainResult}.
The proof of the main result can be divided into the following two steps:
\begin{enumerate}
\item[1.] Construct an $(n,k)$-block code with $\frac{k}{n}= \frac{T-N+1}{T+B-N+1}$ having the following property: The destination can perfectly recover the~$k$ source symbols with decoding delay~$T$ as long as \textcolor{black}{the block code is used over the erasure channel} in~$n$ consecutive time slots.
\item[2.] Convert the $(n,k)$-block code into a convolutional code by periodic interleaving~\cite{Forney1971}.
\end{enumerate}
The details of the above two steps can be found in Section~\ref{sectionMainResultProof} and Section~\ref{sectionPrelim}.

In addition, our construction of optimal streaming codes implies the full characterization of the maximum achievable rate for convolutional codes with any given column distance, column span and decoding delay, whose details can be found in Section~\ref{sectionColumnDistanceSpan}. Simulation results in Section~\ref{sectionNumerical} reveal that our proposed codes outperform all existing practical streaming codes over some instances of the GE channel and the Fritchman channel.
\subsection{Paper Outline}
This paper is organized as follows. The notation in this paper is explained in the next subsection. Section~\ref{sectionDefinition} presents the formulation of streaming codes for the packet erasure channel and states the main result. Section~\ref{sectionPrelim} presents the preliminary results --- a standard procedure for interleaving a block code into a streaming code and two key lemmas which enable us to construct block codes that can be interleaved to form optimal streaming codes. Section~\ref{sectionMainResultProof} contains the proof of the main result, i.e., the existence of optimal streaming codes over the packet erasure channel for all parameters of $(W,T,B,N)$. The optimal streaming codes take the form of convolutional codes obtained by interleaving the block codes as described in the two key lemmas in Section~\ref{sectionPrelim}.
Section~\ref{sectionLemmaHighRate} and Section~\ref{sectionLemmaLowRate} present the proofs of the two key lemmas respectively. In Section~\ref{sectionColumnDistanceSpan}, we discuss the column distance and the column span for low-latency convolutional codes, and use the result in Section~\ref{sectionMainResultProof} to characterize the maximum achievable rate for convolutional codes with fixed column distance, column span and decoding delay. Section~\ref{sectionPracticalImplementation} describes a practical random code construction of optimal low-latency convolutional codes. Section~\ref{sectionNumerical} contains numerical results that compare the performance of the optimal convolutional codes with state-of-the-art schemes over the GE channel and the Fritchman channel. Section~\ref{conclusion} concludes this paper.
\subsection{Notation}\label{notation}
The set of non-negative integers is denoted by $\mathbb{Z}_+$. All the elements of any matrix considered in this paper are taken from a common finite field~$\mathbb{F}$, where~$0$ and~$1$ denote the additive identity and the multiplicative identity respectively. The set of $k$-dimensional row vectors over $\mathbb{F}$ is denoted by $\mathbb{F}^k$, and the set of $k\times n$ matrices over $\mathbb{F}$ is denoted by $\mathbb{F}^{k\times n}$. For any matrix $\mathbf{G}$, we let $\mathbf{G}^t$ and $\rank(\mathbf{G})$ denote respectively the transpose and the rank of $\mathbf{G}$. A row vector in $\mathbb{F}^k$ is denoted by $\mathbf{a} \triangleq [a_0\ a_1\ \ldots \ a_{k-1}]$ where $a_\ell$ denotes the $(\ell+1)^{\text{th}}$ element of $\mathbf{a}$.
The $k$-dimensional identity matrix is denoted by $\mathbf{I}_k$
and the $L\times B$ all-zero matrix is denoted by $\mathbf{0}^{L\times B}$.  An $L \times B$ parity matrix of a systematic maximum-distance separable (MDS) $(L+B, L)$-code is denoted by
$\mathbf{V}^{L\times B}$,
which possesses the property that any $L$ columns of  $[\mathbf{I}_{L} \ \mathbf{V}^{L\times B}]\in \mathbb{F}^{L\times (L+B)}$ are independent. It is well known that a systematic maximum-distance separable (MDS) $(L+B, L)$-code always exists as long as $|\mathbb{F}|\ge L+B$~\cite{MacWilliamsSloane1988}.
 For a matrix $\mathbf{G}\in\mathbb{F}^{k\times n}$, the column space of $\mathbf{G}$ is the set
\textcolor{black}{$
\textrm{space}(\mathbf{G})\triangleq \left\{ \mathbf{G}\boldsymbol{\alpha} \left|\,\boldsymbol{\alpha}\in \mathbb{F}^{n\times 1}\right.\right\}$}.
A $W$-dimensional tuple is denoted by $e^{W}\triangleq(e_0, e_1, \ldots, e_{W-1})$ where $e_i$ denotes the $(i+1)^{\text{th}}$ element of $e^W$. The $W$-dimensional diagonal matrix with diagonal elements $e^W$ is denoted by
\begin{align*}
\diag(e_0,e_1,\ldots, e_{W-1})\triangleq\left[\begin{matrix}e_0 & 0 & \cdots& 0\\0 & e_1 &\cdots & 0\\ \vdots&\vdots & \ddots&\vdots \\0 & \cdots& 0& e_{W-1}\end{matrix}\right].
\end{align*}

\section{Streaming Codes for Channels with Burst and Arbitrary Erasures} \label{sectionDefinition}
This section formally defines our system model, which was briefly described in Section~\ref{subsecChannelModel} and~\ref{subsecContribution}, and states the main result.
\subsection{Problem formulation} \label{subsecFormulation}
The source wants to send a sequence of length-$k$ packets~$\{\mathbf{s}_i\}_{i=0}^{\infty}$ to the destination. Each $\mathbf{s}_i$ is an element in $\mathbb{F}^k$ where $\mathbb{F}$ is some finite field. In each time slot~$i\in\mathbb{Z}_+$, the source packet $\mathbf{s}_i$ is encoded into a length-$n$ packet $\mathbf{x}_i\in\mathbb{F}^n$ to be transmitted to the destination through an erasure channel, and the destination receives $\mathbf{y}_i\in\mathbb{F}^n \cup \{*\}$ where $\mathbf{y}_i$ equals either $\mathbf{x}_i$ or the erasure symbol~`$*$'.
The code is subject to a delay constraint of~$T$ time slots, meaning that the destination must produce an estimate of $\mathbf{s}_i$, denoted by $\hat{\mathbf{s}}_i$, upon receiving $\mathbf{y}_{i+T}$.
In any sliding window that consists of $W\ge T+1$ consecutive time slots, there exists either one burst erasure with length no longer than~$B$ or multiple arbitrary erasures with total count no larger than~$N$. \textcolor{black}{By the assumptions~\eqref{assumptionWLOG} and~\eqref{WgreaterThanT}, we assume
\begin{align}
W>T \ge B \ge N \ge 1\label{parametersFeasible}
\end{align}
unless specified otherwise.
}

\subsection{Standard definitions}
\textcolor{black}{The formal definition of the streaming code described in the previous subsection is stated as follows.}
\begin{Definition}[\textcolor{black}{\cite[Sec.~II-B]{BPKTA17}}] \label{definitionCode}
An $(n, k, T)_{\mathbb{F}}$-streaming code consists of the following:
\begin{enumerate}
\item A sequence of source packets~$\{\mathbf{s}_i\}_{i=0}^{\infty}$ where $\mathbf{s}_i\in \mathbb{F}^k$.
\item An encoding function~$f_i: \underbrace{\mathbb{F}^k \times \ldots \times \mathbb{F}^k}_{i +1 \text{ times }} \rightarrow \mathbb{F}^n$ for each $i\in\mathbb{Z}_+$, where $f_i$ is used by the source at time~$i$ to encode $\mathbf{s}_i$ according to
    \begin{align*}
    \mathbf{x}_i = f_i(\mathbf{s}_0, \mathbf{s}_1, \ldots, \mathbf{s}_i).
    \end{align*}
    \item A decoding function~$\varphi_{i+T}: \underbrace{\mathbb{F}^n\cup\{*\} \times \ldots \times \mathbb{F}^n\cup\{*\}}_{i+T+1 \text{ times }}\rightarrow \mathbb{F}^k$ for each $i\in\mathbb{Z}_+$, where $\varphi_{i+T}$ is used by the destination at time $i+T$ to estimate $\mathbf{s}_i$ according to\footnote{\textcolor{black}{Early decoding is not considered in this definition. In practice, early decoding could decrease the average delay of decoding. See~\cite{BKTZA2017} for an implementation of streaming codes where early decoding is permitted. However, the theoretical and simulation results in this paper remain unchanged even if early decoding is permitted because this paper focuses on maximum rather than average decoding delay.}}
        \begin{align}
       \hat{\mathbf{s}}_i = \varphi_{i+T}(\mathbf{y}_0, \mathbf{y}_1, \ldots, \mathbf{y}_{i+T}). \label{defDecoder}
        \end{align}
\end{enumerate}
\end{Definition}
\medskip
\begin{Definition} \label{definitionConvCode}
An $(n, k, m, T)_{\mathbb{F}}$-convolutional code is an $(n, k, T)_{\mathbb{F}}$-streaming code constructed as follows: Let $\mathbf{G}_0^{\text{conv}}, \mathbf{G}_1^{\text{conv}}, \ldots, \mathbf{G}_{m}^{\text{conv}}$ be $m+1$ generator matrices in $\mathbb{F}^{k\times n}$. Then for each $i\in\mathbb{Z}_+$,
\begin{align}
\mathbf{x}_i = \sum_{\ell=0}^{m} \mathbf{s}_{i-\ell}\,\mathbf{G}_\ell^{\text{conv}} \label{xiConvForm}
\end{align}
where $\mathbf{s}_{-1}=\mathbf{s}_{-2}=\ldots = \mathbf{s}_{-m}=\mathbf{0}^{1\times k}$ by convention.
\end{Definition}
\medskip
\begin{Remark}\label{remark1}
For an $(n, k, m, T)_{\mathbb{F}}$-convolutional code, $m$ is commonly referred to as the \emph{encoder memory}   (see, e.g., \cite[Sec.~1.4]{JohannessonZigangirovBook}), and the role of~$T$ specifies the decoding delay associated with the convolutional code (cf.\ \eqref{defDecoder}).
\end{Remark}
\medskip
\begin{Definition} \label{definitionErasureSeq}
An erasure sequence is a binary sequence denoted by $e^\infty\triangleq\{e_i\}_{i=0}^{\infty}$ where
\begin{equation*}
e_i=\mathbf{1}\{\text{erasure occurs at time~$i$}\}.
\end{equation*}
A $(W, B, N)$-erasure sequence is an erasure sequence $e^\infty$ that satisfies the following: For each $i\in\mathbb{Z}_+$ and any window \begin{equation}
\mathcal{W}_i\triangleq \{i, i+1, \ldots, i+W-1\}, \label{defWindow}
\end{equation}
either $N<\sum_{\ell\in\mathcal{W}_i} e_\ell\le B$ holds with all the~$1$'s in $(e_i, e_{i+1},\ldots,  e_{i+W-1})$ occupying consecutive positions or $\sum_{\ell\in\mathcal{W}_i} e_\ell\le N$ holds with no restriction on the positions of $1$'s. In other words, a $(W, B, N)$-erasure sequence introduces either one burst erasure with length no longer than~$B$ or multiple arbitrary erasures with total count no larger than~$N$ in any window $\mathcal{W}_i$, $i\in\mathbb{Z}_+$. The set of $(W,B,N)$-erasure sequences is denoted by $\Omega_{(W,B,N)}^\infty$.
\end{Definition}
\medskip
\begin{figure}[h!]
\centering
\includegraphics[width=3.5in]{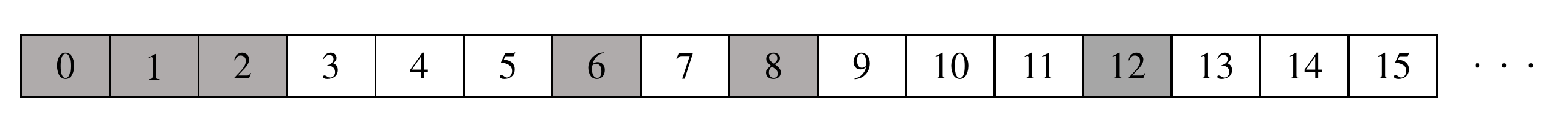}
\caption{A periodic $(5, 3, 2)$-erasure sequence with period $16$.}
\label{figure0}
\end{figure}
\begin{Example}
Suppose $(W,B,N)=(5,3,2)$.
Consider the periodic sequence with period $16$ as shown in Figure~\ref{figure0}. The periodic sequence is in $\Omega_{(5,3,2)}^\infty$ because in any sliding window of length~$W=5$, there is either a single burst erasure of length no longer than $B=3$ or no more than $N=2$ arbitrary erasures.
\end{Example}
\medskip
\begin{Definition} \label{definitionChannel}
The input-output relation of the erasure channel $g_n: \mathbb{F}^n \times \{0,1\} \rightarrow \mathbb{F}^n \cup \{*\}$ subject to~$e\in\{0,1\}$ is defined as
\begin{align}
g_n(\mathbf{x},e)= \begin{cases}\mathbf{x} & \text{if $e=0$,}\\
* & \text{if $e=1$.}
\end{cases} \label{defChannelLaw}
\end{align}
For any erasure sequence $e^\infty$ and any $(n, k, T)_{\mathbb{F}}$-streaming code, the following input-output relation holds for the erasure channel for each $i\in\mathbb{Z}_+$:
\begin{align}
\mathbf{y}_i = g_n(\mathbf{x}_i, e_i).\label{defChannelOutput}
\end{align}
\end{Definition}
\medskip
\begin{Definition} \label{definitionAchievability}
An $(n, k, T)_{\mathbb{F}}$-streaming code is said to be $(W, B, N)$-achievable if the following holds for any $(W, B, N)$-erasure sequence~$e^\infty\in\Omega_{(W,B,N)}^{\infty}$: For all $i\in\mathbb{Z}_+$ and all $\mathbf{s}_i\in \mathbb{F}^k$, we have
\begin{equation*}
 \hat{\mathbf{s}}_i = \mathbf{s}_i
\end{equation*}
where
\begin{align*}
 \hat{\mathbf{s}}_i &=  \textcolor{black}{\varphi_{i+T}\big(\mathbf{y}_0 , \ldots, \mathbf{y}_{i+T}\big)}\\*
 & = \varphi_{i+T}\big( g_n(\mathbf{x}_0, e_0), \ldots,  g_n(\mathbf{x}_{i+T}, e_{i+T})\big)
\end{align*}
due to~\eqref{defDecoder} and~\eqref{defChannelOutput}.
\end{Definition}
\medskip
\begin{Definition} \label{defCapacity}
Fix any $(W,T,B,N)$ that satisfies~\eqref{parametersFeasible}. The $(W, T, B, N)$-capacity, denoted by $C_{(W, T,B,N)}$, is the \textcolor{black}{supremum of the rates attained by} $(n, k, T)_{\mathbb{F}}$-streaming codes that are $(W, B, N)$-achievable, i.e.,
\begin{equation*}
C_{(W, T,B,N)}\triangleq \textcolor{black}{\sup}\left\{\left.\frac{k}{n} \,\right| \text{There exists an $(W, B, N)$-achievable $(n, k, T)_{\mathbb{F}}$-streaming code for some $\mathbb{F}$} \right\}.
\end{equation*}
\end{Definition}
\medskip

\textcolor{black}{It was shown in~\cite[Th.~1 and Th.~2]{BPKTA17} that
\begin{align}
\frac{T-N}{T+B-N} \le C_{(W,T,B,N)} \le \frac{T-N+1}{T+B-N+1} \label{boundsOnCapacity}
\end{align}
holds for any~$(W, T, B, N)$. Our main result stated in the next subsection closes the gap.}

\subsection{Main Result} \label{subsecMainResult}
\begin{Theorem}\label{thmMainResult}
Fix any $(W,T,B,N)$ that satisfies~\eqref{parametersFeasible} and suppose
\begin{align}
\left|\mathbb{F}\right|> 2\left({{T+1}\choose{N}} + T-B+2\right). \label{assumptionFieldSize}
 \end{align}
Then, there exists an $(n, k, T, T)_{\mathbb{F}}$-convolutional code that is $(W,B,N)$-achievable where $k= T-N+1$ and $n= T+B-N+1$.
\end{Theorem}
\medskip

Combining Theorem~\ref{thmMainResult}, Definition~\ref{defCapacity} and~\eqref{boundsOnCapacity},
 we fully characterize the $(W,T,B,N)$-capacity to be
\begin{align*}
C_{(W,T,B,N)} = \frac{T-N+1}{T+B-N+1}
\end{align*}
for all $(W,T,B,N)$ that satisfies~\eqref{parametersFeasible}, which generalizes the capacity results for the special case $N=1$ in~\cite{MartinianSundberg2004} and for the special case $\frac{T-N+1}{T+B-N+1}=\frac{1}{2}$ in~\cite{BKTA2013}. \textcolor{black}{In particular, the upper bound in~\eqref{boundsOnCapacity} obtained in~\cite{BPKTA17} is tight and the supremum in Definition~\ref{defCapacity} can be replaced with a maximum.}
\section{Preliminaries for the Proof of Theorem~\ref{thmMainResult}} \label{sectionPrelim}
\textcolor{black}{An important step of the proof of Theorem~\ref{thmMainResult} is to construct streaming codes by periodically interleaving block codes. The definition of a block code is formally stated as follows.}
\textcolor{black}{\begin{Definition} \label{definitionBlockCode}
An $(n, k, T)_{\mathbb{F}}$-block code consists of the following:
\begin{enumerate}
\item A sequence of $k$ source symbols~$\{s[i]\}_{i=0}^{k-1}$ where $s[i]\in\mathbb{F}$.
\item A generator matrix $\mathbf{G}\in \mathbb{F}^{k\times n}$ defined as
\begin{align*}
\mathbf{G} \triangleq \left[
\begin{matrix}
\mathbf{I}_{k} & \mathbf{P}
\end{matrix}
\right]
\end{align*}
where $\mathbf{P}\in \mathbb{F}^{k\times (n-k)}$ is some parity-check matrix to be determined later. The codeword is generated according to
\begin{align}
\big[x[0]\ x[1]\ \ldots \ x[n-1]]=[s[0]\ s[1]\ \ldots \ s[k-1]\big]\, \mathbf{G}. \label{encodingOfBlockCode}
\end{align}
    \item A decoding function~$\varphi_{i+T}: \underbrace{\mathbb{F}\cup\{*\} \times \ldots \times \mathbb{F}\cup\{*\}}_{i+T+1 \text{ times }}\rightarrow \mathbb{F}$ for each $i\in\{0, 1, \ldots, k-1\}$, where $\varphi_{i+T}$ is used by the destination at time $i+T$ to estimate $s[i]$ according to
        \begin{align*}
       \hat{s}[i] =
       \begin{cases}
        \varphi_{i+T}(y[0] ,y[1],\ldots, y[i+T]) & \text{if $i+T\le n-1$,} \\
        \varphi_{i+T}(\underbrace{y[0] ,y[1],\ldots, y[n-1], * , \ldots, *}_{i+T+1\text{ symbols}})&  \text{if $i+T> n-1$}.
        \end{cases}
        \end{align*}
\end{enumerate}
\end{Definition}}

\textcolor{black}{The following definition concerns the error-correcting capability of $(n, k, T)_{\mathbb{F}}$-block codes.}
\medskip
\begin{Definition} \label{definitionAchievabilityBlock}
 An $(n, k, T)_{\mathbb{F}}$-block code is said to be $(W, B, N)$-achievable if the following holds for any $(W, B, N)$-erasure sequence~$e^\infty\in \Omega_{(W,B,N)}^\infty$\footnote{\textcolor{black}{Only the first~$n$ elements of $e^\infty$ play a role in the definition.}}: For the $(n, k, T)_{\mathbb{F}}$-block code, we have
\begin{equation*}
\textcolor{black}{ \hat{s}[i] = s[i]}
\end{equation*}
for all $i\in\{0, 1, \ldots, k-1\}$ and all $s[i]\in \mathbb{F}$, where
\textcolor{black}{ \begin{align}
       \hat{s}[i] =
       \begin{cases}
        \varphi_{i+T}(g_1(x[0], e_0),\ldots, g_1(x[i+T], e_{i+T})) & \text{if $i+T\le n-1$,} \\
        \varphi_{i+T}(\underbrace{g_1(x[0], e_0),\ldots, g_1(x[n-1], e_{n-1}), * , \ldots, *}_{i+T+1\text{ symbols}})&  \text{if $i+T> n-1$}
        \end{cases} \label{defChannelLawBlock}
        \end{align}}
       with $g_1$ being defined in~\eqref{defChannelLaw}.
\end{Definition}
\medskip

\textcolor{black}{The following lemma implies that constructing a $(W, B, N)$-achievable convolutional code is not more difficult than constructing a $(W, B, N)$-achievable block code.} The proof of the following lemma is deferred to Appendix~\ref{appendixA} because it follows the standard argument of interleaving a block code into a convolutional code by means of periodic interleaving~\cite{Forney1971} (see also \cite[Sec.\ IV-A]{MartinianSundberg2004}).
\begin{Lemma}\label{lemmaBlockToStreaming} 
Given an $(n, k, T)_{\mathbb{F}}$-block code which is $(W, B, N)$-achievable, we can construct an $(n, k, n-1, T)_{\mathbb{F}}$-convolutional code which is $(W, B, N)$-achievable. More specifically, given that $
\mathbf{G}= \big[g_{i,j}\big]_{\substack{0\le i\le k-1,\\ 0\le j\le n-1}} %
$
is the generator matrix of the $(n, k, T)_{\mathbb{F}}$-block code where $g_{i,j}$ is the entry situated in row~$i$ and column~$j$ of~$\mathbf{G}$, we can construct the~$n-1$ generator matrices of the $(n, k, n-1, T)_{\mathbb{F}}$-convolutional code as follows: For each $\ell\in\{0, 1, \ldots, n-1\}$, construct
\begin{align}
\mathbf{G}_\ell^{\text{conv}} \triangleq
\begin{cases}
\left[\mathbf{0}^{k\times \ell}\ \diag\big(g_{0,\ell},g_{1,\ell+1}, \ldots, g_{k-1,\ell+k-1}  \big)\ \mathbf{0}^{k\times (n-k-\ell)}\right]  & \text{if $0\le \ell \le n-k$,} \\
\left[\begin{array}{c:c}\mathbf{0}^{k\times \ell} & \begin{matrix}\diag\big(g_{0,\ell},g_{1,\ell+1}, \ldots, g_{n-1-\ell,n-1}  \big)\\ \mathbf{0}^{(k-n+\ell)\times (n-\ell)}\end{matrix}\end{array}\right] & \text{if $n-k<\ell \le n-1$,}
\end{cases}
 \label{appendixAeq3}
\end{align}
where $\mathbf{G}=\sum_{\ell=0}^{n-1}\mathbf{G}_\ell^{\text{conv}}$.
\textcolor{black}{In particular, if we let
$\mathbf{s}_i \triangleq [s_i[0]\ s_i[1]\ \cdots \ s_i[k-1] ] 
$
and let \begin{align*}
\big[x_i[0]\ x_{i+1}[1]\ \cdots \ x_{i+n-1}[n-1] \big] &\triangleq \big[s_i[0]\ s_{i+1}[1]\ \cdots \ s_{i+k-1}[k-1] \big] \mathbf{G} 
\end{align*}
for all $i\in\mathbb{Z}_+$, then the symbols generated at time~$i$ by the $(n, k, n-1, T)_{\mathbb{F}}$-convolutional code are
\begin{align}
\mathbf{x}_i 
\triangleq\big[x_i[0]\ x_{i}[1]\ \cdots \ x_{i}[n-1] \big]. \label{st2LemmaBlockToStreaming}
\end{align}}
\end{Lemma}

\begin{Example}\label{exampleBlockToConv}
Suppose we are given a $(5, 3, 2)$-achievable $(6, 3, 4)_{\mathbb{F}}$-block code with generator matrix
\begin{align*}
\mathbf{G}=\left[\begin{array}{cccccc}1&0&0& 1&1&0\\0&1&0&0&1&1\\0&0&1&0&1&2 \end{array}\right].
\end{align*}
 Let $\{\mathbf{s}_i\}_{i\in\mathbb{Z}_+}$ be a streaming message where \textcolor{black}{$\mathbf{s}_i=\big[s_i[0] \ s_i[1] \ s_i[2]\big]\in \mathbb{F}^{3}$}. From time~$i-2$ to $i+5$, the symbols yielded by the $(6, 3, 5, 4)_{\mathbb{F}}$-convolutional code constructed by interleaving the $(6, 3, 4)_{\mathbb{F}}$-block code according Lemma~\ref{lemmaBlockToStreaming} are shown in Table~\ref{table1}.
\begin{table}[!t]
\centering
\begin{tabular}{|c|*{8}{c|}}\hline
\backslashbox{\small Symbol}{\small Time}
&\makebox[1.5em]{$i-2$}&\makebox[1.5em]{$i-1$}&\makebox[1.5em]{$i$}&\makebox[1.5em]{$i+1$}&\makebox[1.5em]{$i+2$}
&\makebox[1.5em]{$i+3$}&\makebox[1.5em]{$i+4$}&\makebox[1.5em]{$i+5$}\\\hline
$0$ &\textcolor{red}{$s_{i-2}[0]$} &\magenta{$s_{i-1}[0]$} & \blue{$s_i[0]$} &$s_{i+1}[0]$&$s_{i+2}[0]$&$s_{i+3}[0]$ &$s_{i+4}[0]$&$s_{i+5}[0]$\\\hline
$1$ & $s_{i-2}[1]$ &$\textcolor{red}{s_{i-1}[1]}$ & $\magenta{s_i[1]}$ & \blue{$s_{i+1}[1]$}&$s_{i+2}[1]$&$s_{i+3}[1]$ &$s_{i+4}[1]$&$s_{i+5}[1]$\\\hline
$2$ &$s_{i-2}[2]$ &$s_{i-1}[2]$ & $\textcolor{red}{s_i[2]}$ &\magenta{$s_{i+1}[2]$}& \blue{$s_{i+2}[2]$}&$s_{i+3}[2]$ &$s_{i+4}[2]$&$s_{i+5}[2]$\\\hline
$3$ &$\ddots$ & $\ddots$& $\ddots$ &\textcolor{red}{$s_{i-2}[0]$} &\magenta{$s_{i-1}[0]$} &  \blue{$s_i[0]$} &$\ddots$&$\ddots$\\\hline
$4$ &$\ddots$ &$\ddots$ &$\ddots$  &$\ddots$ &\parbox[c]{0.6 in}{\textcolor{red}{$s_{i-2}[0]\\ +s_{i-1}[1]\\+ s_{i}[2]$}} &  \parbox[c]{0.6 in}{\magenta{$s_{i-1}[0]\\ +s_{i}[1]\\+ s_{i+1}[2]$}}&   \parbox[c]{0.6 in}{\blue{$s_i[0] \\+s_{i+1}[1]\\+ s_{i+2}[2]$}} &$\ddots$\\\hline
$5$ &$\ddots$ &$\ddots$ & $\ddots$ &$\ddots$&$\ddots$&\parbox[c]{0.6 in}{\textcolor{red}{$s_{i-1}[1]\\+ 2s_{i}[2]$}}&\parbox[c]{0.6 in}{\magenta{$s_{i}[1]\\+ 2s_{i+1}[2]$}}&\parbox[c]{0.6 in}{\blue{$s_{i+1}[1]\\+ 2s_{i+2}[2]$}} \\\hline
\end{tabular}\smallskip
\caption{Symbols yielded by a $(6, 3, 5, 4)_{\mathbb{F}}$-convolutional code through interleaving a $(6, 3, 4)_{\mathbb{F}}$-block code.}
\label{table1}
\end{table}
 The symbols in Table~\ref{table1} which are highlighted in the same color diagonally (in~$\searrow$ direction) are encoded using the same $(5, 3, 2)$-achievable $(6, 3, 4)_{\mathbb{F}}$-block code. Given the fact that each $(6, 3, 4)_{\mathbb{F}}$-block code is $(5, 3, 2)$-achievable, we can see from Table~\ref{table1} that \textcolor{black}{$\mathbf{s}_i=\big[s_i[0] \ s_i[1] \ s_i[2]\big]$} can be perfectly recovered by time $i+5$ as long as the erasure sequence is taken from $\Omega_{(5,3,2)}^\infty$.
\end{Example}
\medskip

\textcolor{black}{Instead of proving Theorem~\ref{thmMainResult} by enumerating all possible $(W, B, N)$-erasure sequences, we will prove Theorem~\ref{thmMainResult} by enumerating a small subset of sequences called \textit{maximal $(W, B, N)$-erasure patterns}.}
\begin{Definition}\label{definitionMaximalErasure}
A maximal $(W, B, N)$-erasure pattern is a $W$-dimensional binary tuple $\varepsilon^{W}$ that satisfies either $\sum_{\ell=0}^{W-1} \varepsilon_\ell=B$ with all the $1$'s in $\varepsilon^{W}$ occupying consecutive positions or $\sum_{\ell=0}^{W-1} \varepsilon_\ell= N$ with no restriction on the positions of $1$'s. The set of maximal $(W, B, N)$-erasure patterns is denoted by $\Omega_{B,N}^{W}$.
\end{Definition}

Recall the definition of window $\mathcal{W}_i$ in~\eqref{defWindow} (where $|\mathcal{W}_i|=W$). For any $(W, B, N)$-erasure sequence $e^\infty$ and any $\mathcal{W}_i$, there always exists a maximal $(W, B, N)$-erasure pattern $\varepsilon^{W}$ such that $e_\ell \le \varepsilon_\ell$ for all $\ell\in\mathcal{W}_i$ by Definition~\ref{definitionErasureSeq} and Definition~\ref{definitionMaximalErasure}. The following lemma enables us to prove Theorem~\ref{thmMainResult} by considering only maximal $(T+1, B, N)$-erasure patterns in $\Omega_{B,N}^{T+1}$ rather than all possible $(W, B, N)$-erasure sequences in $\Omega_{(W,B,N)}^{\infty}$. Before presenting the lemma, we define the following notations which will be used in the rest of the paper. We let $\mathbf{u}_i^{(k)}$ denote the $k$-dimensional unit column vector $[\mathbf{0}^{1\times i} \ 1 \  \mathbf{0}^{1\times (k-i-1)}]^t$ for each $i\in\{0, 1, \ldots, k-1\}$, let
\begin{align}
\mathbf{I}_{j}^{(k)}\triangleq\left[\begin{array}{ll} \mathbf{0}^{(k-j)\times (k-j)} &  \mathbf{0}^{(k-j)\times j}\\ \mathbf{0}^{j\times (k-j)} & \mathbf{I}_{j}\end{array}\right] \label{defMatrixI}
\end{align}
be the $k\times k$ diagonal matrix which embeds~$\mathbf{I}_{j}$ as a submatrix for each $j\in\{0,1, \ldots, k\}$, and let
\begin{align}
\mathbf{E}_{\varepsilon^j}\triangleq \mathbf{I}_{j}-\diag(\varepsilon^j) \label{defMatrixE}
\end{align}
be the~$j\times j$ diagonal matrix with diagonal elements $(1-\varepsilon_0), (1-\varepsilon_1), \ldots, (1-\varepsilon_{j-1})$ for any length-$j$ binary tuple $\varepsilon^j$. We will always multiply $\mathbf{E}_{\varepsilon^j}$ on the right side of a matrix having $j$ columns, and the multiplication characterizes the erasure operation introduced by~$\mathbf{E}_{\varepsilon^j}$ by zeroing the columns of the multiplied matrix indicated by~$\varepsilon^j$.
The proof of the following lemma is straightforward and hence relegated to Appendix~\ref{appendixB}.
\medskip
\begin{Lemma} \label{lemmaSufficientCondition}
Fix any $(W,T,B,N)$ that satisfies~\eqref{parametersFeasible}. Let $\mathbf{G} = [\mathbf{g}_0\ \mathbf{g}_1\ \ldots \ \mathbf{g}_{n-1}]$ be a $k\times n$  matrix in $\mathbb{F}^{k\times n}$, and let
\begin{align}
\mathbf{G}_i\triangleq
\begin{cases}
[\mathbf{g}_i\ \mathbf{g}_{i+1}\ \ldots \ \mathbf{g}_{i+T}]&\text{if $i\le n-T-1$,}\\
\left[\begin{array}{cccc:c}\mathbf{g}_i & \mathbf{g}_{i+1}& \ldots & \mathbf{g}_{n-1} & \mathbf{0}^{k\times (i+T-n+1)}\end{array}\right] &   \text{if $i> n-T-1$}
\end{cases} \label{defMatrixGi}
\end{align}
be a submatrix of $\mathbf{G}$ for each $i\in\{0, 1, \ldots, k-1\}$.
There exists an $(n, k, T)_{\mathbb{F}}$-block code with generator matrix~$\mathbf{G}$ which is $(W, B, N)$-achievable if $\mathbf{G}$ satisfies the following sufficient condition:
\\ \indent
  For each~$i\in\{0, 1, \ldots, k-1\}$ and each maximal $(T+1,B,N)$-erasure pattern $\varepsilon^{T+1}\in \Omega_{B,N}^{T+1}$, it is true that
    \begin{equation}
    \mathbf{u}_i^{(k)} \in \textrm{space}\left(\mathbf{I}_{k-i}^{(k)} \,\mathbf{G}_i \mathbf{E}_{\varepsilon^{T+1}}\right). \label{stLemmaSufficientCondition}
    \end{equation}
\end{Lemma}
\medskip
\begin{Remark}\label{remark2}
Lemma~\ref{lemmaSufficientCondition} transforms the problem of finding optimal $(n, k, T)_{\mathbb{F}}$-block codes that are $(W,B,N)$-achievable into a purely algebraic problem stated in~\eqref{stLemmaSufficientCondition}. The physical meaning of~\eqref{stLemmaSufficientCondition} can be interpreted as follows: Suppose $[x[0]\ x[1]\ \cdots \ x[n-1]]= [s[0]\ s[1]\ \cdots \ s[k-1]]\,\mathbf{G}$. Then, \eqref{stLemmaSufficientCondition} implies that
$s_i$ can be perfectly recovered by time $i+T$ as long as $s_0, s_1, \ldots, s_{i-1}$ have been perfectly recovered and the erasure patten in $\mathcal{W}_i$ is in $ \Omega_{B,N}^{T+1}$.
\end{Remark}
\medskip

The following lemma shows the existence of a generator matrix $\mathbf{G}$ which satisfies the sufficient condition in~\eqref{stLemmaSufficientCondition} when $T-N+1 \ge B$. One component of the generator matrix is an \emph{$m \times (N+m)$ $N$-diagonal matrix} defined as
\begin{align}
\mathbf{D}_N^{m\times (N+m)} \triangleq \left[\begin{array}{ccccccc} d_0^{(0)} &\cdots &d_{N-1}^{(0)} & 0&\cdots  &\cdots&0 \\ 0 & d_0^{(1)} & \cdots & d_{N-1}^{(1)} & 0&\cdots & 0\\\vdots &\ddots & \ddots&  \ddots& \ddots & \ddots &\vdots\\  0 & \cdots & 0 & d_0^{(m-1)} & \ldots & d_{N-1}^{(m-1)}& 0\end{array}\right] \label{defMatrixD}
\end{align}
with arbitrary values for $\{d_\ell^{(i)}\}_{\substack{0\le i\le m-1 \\ 0\le \ell \le N-1 }}$.
The proof of the following lemma is tedious and is therefore deferred to Section~\ref{sectionLemmaHighRate}.

\medskip
\begin{Lemma} \label{lemmaHighRate}
Fix any $(W,T,B,N)$ that satisfies~\eqref{parametersFeasible} and let $k\triangleq T-N+1$ and $n\triangleq k+B$. Suppose $k\ge B$, which is equivalent to $k/n \ge 1/2$ (high-rate regime). If $\mathbb{F}$ satisfies~\eqref{assumptionFieldSize}, there exists a $\mathbf{P}$ having the form
\begin{align}
\mathbf{P}\triangleq\left[\begin{array}{c:c} \multicolumn{2}{c}{\mathbf{D}_N^{(B-N)\times B}} \\\hdashline  \begin{array}{l}\mathbf{0}^{N \times (B-N)} \end{array}& \mathbf{P}_{\text{right}}\\ \hdashline \multicolumn{2}{c}{\mathbf{V}^{(k-B)\times B}} \end{array}\right] \label{parityHighRate}
\end{align}
such that $\mathbf{G} = [\,\mathbf{I}_k \ \mathbf{P}]$ satisfies~\eqref{stLemmaSufficientCondition} for all $i\in\{0, 1, \ldots, k\}$ and all $\varepsilon^{T+1}\in \Omega_{B,N}^{T+1}$, where $\mathbf{D}_N^{m\times (N+m)}$ is an $N$-diagonal matrix as defined in~\eqref{defMatrixD}, $\mathbf{P}_{\text{right}}$ is a $N\times N$ matrix with non-zero entries, and $\mathbf{V}^{(k-B)\times B}$ denotes a $(k-B)\times B$ parity matrix of a systematic MDS code.
\end{Lemma}
\medskip
\begin{Remark}\label{remark3}
For the special case $N=1$ with delay $T=k+N-1=k$, the parity-check matrix $\mathbf{P}$ in Lemma~\ref{lemmaHighRate} reduces to the parity-check matrix of the Martinian-Sundberg scheme~\cite[Th.~2]{MartinianSundberg2004} in which $\mathbf{P}$ was simply chosen to be $\left[\begin{matrix}\mathbf{I}_{B}\\ \mathbf{V}^{(k-B)\times B}\end{matrix}\right] $. For the case $N>1$ with delay $T=k+N-1>k$, the Martinian-Sundberg scheme is no longer $(W,B,N)$-achievable because the row weight (number of non-zero elements) in each of the first~$B$ columns in the generator matrix of the base block code equals~$2$, implying that the contribution of some source symbol can be completely erased by some choice of~$2$ arbitrary erasures. In contrast, our choice of $\mathbf{P}$ in Lemma~\ref{lemmaHighRate} having the form~\eqref{parityHighRate} ensures that the minimum row weight of the generator matrix is~$N+1$, implying that the contribution of every source symbol is not completely erased by any choice of~$N$ arbitrary erasures. Since $T=k+N-1$ and $n=k+B$, it follows that $B-N$ symbols encoded by $\mathbf{G}$ need to be decoded before the whole block has been received.
\end{Remark}
\medskip
\begin{Remark}\label{remark4}
For the special case $N=B$ with delay $T=k+N-1 = n-1$, we can simply choose $\mathbf{P}$ in Lemma~\ref{lemmaHighRate} to be $\mathbf{V}^{k\times B}$ such that the resultant code is a maximum distance separable (MDS) code. In this case, the decoding of every symbol encoded by $\mathbf{G}$ can be performed after the whole block has been received because $T= n-1$.
\end{Remark}
\medskip
\begin{Example}\label{example1}
Suppose $(W, T, B,N)=(6, 5,3, 2)$ where $k=4 \ge B$. Fix $\mathbb{F}=\mathrm{GF}(41)$ so that \eqref{assumptionFieldSize} is satisfied. By Lemma~\ref{lemmaHighRate}, there exists a $\mathbf{G} = [\,\mathbf{I}_k \ \mathbf{P}]$ with $\mathbf{P}$ having the form~\eqref{parityHighRate} such that $\mathbf{G}$  satisfies~\eqref{stLemmaSufficientCondition}.
A candidate for such a~$\mathbf{G}$ is
\begin{align*}
\mathbf{G}=\left[\begin{array}{c} \begin{array}{cccc:ccc}1&0&0&0&1&2&0\\0&1&0&0&0&1&3\\0&0&1&0&0&2&1\\0&0&0&1&1&1&1 \end{array}\end{array}\right],
\end{align*}
where the minimum row weight of~$\mathbf{G}$ equals~$3$. \textcolor{black}{In particular, condition~\eqref{stLemmaSufficientCondition} is satisfied for each $i\in\{0, 1, 2, 3\}$ and each maximal $(6,3,2)$-erasure pattern $\varepsilon^{6}\in \Omega_{(6,3,2)}^{6}$  due to the following two facts:
\begin{align*}
\left[\begin{array}{c}1\\0\\0\\0 \end{array}\right] \in \textrm{space}\left(\left[\begin{array}{c} \begin{array}{cccc:cc}1&0&0&0&1&2\\0&1&0&0&0&1\\0&0&1&0&0&2\\0&0&0&1&1&1 \end{array}\end{array}\right] \mathbf{E}_{\varepsilon^6}\right)
\end{align*}
and
\begin{align*}
\textrm{space}\left(\left[\begin{array}{ccc} 0&0&0\\ 1&0&0\\ 0&1&0\\ 0&0&1 \end{array}\right]\right) \subseteq \textrm{space}\left(\left[\begin{array}{c} \begin{array}{ccc:ccc}0&0&0&0&0&0\\1&0&0&0&1&3\\0&1&0&0&2&1\\0&0&1&1&1&1 \end{array}\end{array}\right]\mathbf{E}_{\varepsilon^6}\right)
\end{align*}
where $\mathbf{E}_{\varepsilon^6}=\diag(1-e_0,1-e_1,\ldots, 1-e_{5})$. The effect of $\mathbf{E}_{\varepsilon^6}$ is to replace the columns of the multiplied matrix whose indices are inside $\left\{i\in\{0, 1, 2,3,4, 5\}|e_i=1\right\}$ with $\mathbf{0}^{4\times 1}$, which is equivalent to ``erasing" those columns when we evaluate the column space of the multiplied matrix. Since $\varepsilon^{6}$ is an arbitrary maximal $(6,3,2)$-erasure pattern, the erased columns specified by~$\mathbf{E}_{\varepsilon^6}$ take the form of any consecutive~$3$ columns or any~$2$ arbitrary columns. The intuition behind the idea of finding~$\mathbf{G}$ is explained as follows. Consider the baseline Martinian-Sundberg matrix (cf.\ Remark~\ref{remark3}) denoted by
\begin{align*}
\mathbf{G}^* \triangleq \left[\begin{array}{c} \begin{array}{cccc:ccc}1&0&0&0&1&0&0\\0&1&0&0&0&1&0\\0&0&1&0&0&0&1\\0&0&0&1&1&1&1 \end{array}\end{array}\right]
\end{align*}
 which has the same dimension as the desired~$\mathbf{G}$ with dimension $(T-N+1)\times (T+B-N+1)=4\times 7$. Since the minimum row weight of~$\mathbf{G}^*$ equals~$2$, some symbols cannot be recovered if the channel is subject to~$2$ arbitrary erasures. Therefore, we are motivated to construct a~$\mathbf{G}$ with minimum row weight~$3$ by replacing some zeros in~$\mathbf{G}^*$ with non-zeros so that~$\mathbf{G}$ would satisfy~\eqref{stLemmaSufficientCondition} for each $i\in\{0, 1, 2, 3\}$.
}
This example remains valid if we replace $\mathrm{GF}(41)$ by $\mathrm{GF}(5)$, which is not surprising because~\eqref{assumptionFieldSize} is only a sufficient condition on $\mathbb{F}$.
\end{Example}
\medskip
\begin{Example}\label{example1*}
Suppose $(W, T, B,N)=(8, 7,4, 2)$ where $k=6 \ge B$. Fix $\mathbb{F}=\mathrm{GF}(67)$ so that \eqref{assumptionFieldSize} is satisfied. By Lemma~\ref{lemmaHighRate}, there exists
a $\mathbf{G} = [\,\mathbf{I}_k \ \mathbf{P}]$ with $\mathbf{P}$ having the form~\eqref{parityHighRate} such that $\mathbf{G}$  satisfies~\eqref{stLemmaSufficientCondition}.
An example for such a~$\mathbf{G}$ is
\begin{align*}
\mathbf{G}=\left[
\begin{array}{cccccc:cccc}1 &0 &0 &0 & 0&0 &1 &6& 0 &0\\0 &1&0&0&0&0&0 &5 &5^2 &0 \\0&0&1 &0&0&0 &0&0 &4^2 &4^3\\0&0& 0&1&0&0&0 &0 &3^2 &3^3 \\0&0& 0&0&1&0&1 &2 &2^2 &2^3 \\0&0& 0&0&0&1&1 &1 &1 &1\end{array}
\right].
\end{align*}
\end{Example}

The following lemma shows the existence of a generator matrix $\mathbf{G}$ which satisfies the sufficient condition in~\eqref{stLemmaSufficientCondition} when $T-N+1 < B$.
The proof is tedious and is therefore deferred to Section~\ref{sectionLemmaLowRate}.
\medskip
\begin{Lemma}\label{lemmaLowRate}
Fix any $(W,T,B,N)$ that satisfies~\eqref{parametersFeasible} and let $k\triangleq T-N+1$ and $n\triangleq k+B$. Suppose $k< B$, which is equivalent to $k/n < 1/2$ (low-rate regime. If $\mathbb{F}$ satisfies~\eqref{assumptionFieldSize}, there exists a $\mathbf{P}$ having the form
\begin{align}
\mathbf{P}\triangleq
\left[\begin{array}{c:c} \begin{matrix}\mathbf{P}_{\text{left}}\\ \mathbf{V}^{(k-B+N)\times (B-k)}\end{matrix}  & \begin{array}{c:c} \multicolumn{2}{c}{\mathbf{D}_{k-B+N}^{(B-N)\times k}} \\ \hdashline \begin{array}{l}\ \mathbf{0}^{(k-B+N) \times (B-N)} \end{array}& \begin{matrix} \mathbf{V}^{(k-B+N)\times (k-B+N)}\end{matrix}\end{array} \end{array}\right] \label{parityLowRate}
\end{align}
such that $\mathbf{G} = [\,\mathbf{I}_k \ \mathbf{P}]$ satisfies~\eqref{stLemmaSufficientCondition}  for all $i\in\{0, 1, \ldots, k\}$ and all $\varepsilon^{T+1}\in \Omega_{B,N}^{T+1}$, where $\mathbf{P}_{\text{left}}$ is a $(B-N)\times (B-k)$ matrix, $\mathbf{D}_{k-B+N}^{(B-N)\times k}$ is a $(k-B+N)$-diagonal matrix as defined in~\eqref{defMatrixD}, and
 $\mathbf{V}^{(k-B+N)\times N} \triangleq\left[\mathbf{V}^{(k-B+N)\times (B-k)} \ \mathbf{V}^{(k-B+N)\times (k-B+N)}\right]$ constitutes a $(k-B+N)\times N$ parity matrix of a systematic MDS code.
\end{Lemma}
\medskip
\begin{Remark}\label{remark5}
Suppose $k<B$. Then $N>1$ must hold, and our choice of $\mathbf{P}$ in Lemma~\ref{lemmaLowRate} having the form~\eqref{parityLowRate} ensures that the minimum row weight of the generator matrix is~$N+1$. As in the case $k\ge B$ discussed in Remark~\ref{remark3}, we see from~\eqref{parityLowRate} that the contribution of every source symbol is not completely erased by any choice of~$N$ arbitrary erasures, and $B-N$ symbols encoded by $\mathbf{G}$ need to be decoded before the whole block has been received.
\end{Remark}
\medskip
\begin{Example}\label{example2}
Suppose $(W, T, B,N)=(6, 5,4, 3)$ where $k=3<B$. Fix $\mathbb{F}=\mathrm{GF}(47)$ so that \eqref{assumptionFieldSize} is satisfied. By Lemma~\ref{lemmaLowRate}, there exists a $\mathbf{G} = [\,\mathbf{I}_k \ \mathbf{P}]$ with $\mathbf{P}$ having the form~\eqref{parityLowRate} such that $\mathbf{G}$ satisfies~\eqref{stLemmaSufficientCondition}.
A candidate for such a~$\mathbf{G}$ is
\begin{align*}
\mathbf{G}=\left[\begin{array}{c} \begin{array}{ccc:cccc}1&0&0&1&1&3&0\\0&1&0&1&0&2&4\\0&0&1&1&0&1&1 \end{array}\end{array}\right],
\end{align*}
where the minimum row weight of~$\mathbf{G}$ equals~$4$.
This example remains valid if we replace $\mathrm{GF}(47)$ by $\mathrm{GF}(5)$, which is not surprising because~\eqref{assumptionFieldSize} is only a sufficient condition on $\mathbb{F}$.
\end{Example}
\medskip
\begin{Example}\label{example3}
Suppose $(W, T, B,N)=(8, 7,6,4)$ where $k=4<B$. Fix $\mathbb{F}=\mathrm{GF}(149)$ so that \eqref{assumptionFieldSize} is satisfied. By Lemma~\ref{lemmaLowRate}, there exists a $\mathbf{G} = [\,\mathbf{I}_k \ \mathbf{P}]$ with $\mathbf{P}$ having the form~\eqref{parityLowRate} such that $\mathbf{G}$ satisfies~\eqref{stLemmaSufficientCondition}.
A candidate for such a $\mathbf{G}$ is
\begin{align}\mathbf{G}=
\left[
\begin{array}{cccc:cccccc}1 &0 &0 &0&1&4&4^2&4^3&0& 0 \\0 &1&0&0&1&3&0&3^3&3^4&0 \\0&0&1 &0&1&2&0&0&2^4  &2^5 \\ 0&0&0 &1&1&1&0&0&1  &1\end{array}
\right].
\end{align}
\end{Example}
\section{Proof of Theorem~\ref{thmMainResult}} \label{sectionMainResultProof}
Fix any $(W,T,B,N)$ that satisfies~\eqref{parametersFeasible} and choose a sufficiently large $\mathbb{F}$ which satisfies~\eqref{assumptionFieldSize}. Let $k\triangleq T-N+1$ and $n\triangleq k+B$. Consider the following two cases:
\\
\textbf{Case $k\ge B$:}\\ \indent
By Lemma~\ref{lemmaHighRate} and Lemma~\ref{lemmaSufficientCondition}, there exists an $(n, k, T)_{\mathbb{F}}$-block code with generator matrix~$\mathbf{G}=[\,\mathbf{I}_k \ \mathbf{P}]\in \mathbb{F}^{k\times n}$ which is $(W, B, N)$-achievable where $\mathbf{P}$ has the form~\eqref{parityHighRate}.
\\
\textbf{Case $k<B$:}\\ \indent
By Lemma~\ref{lemmaLowRate} and Lemma~\ref{lemmaSufficientCondition}, there exists an $(n, k, T)_{\mathbb{F}}$-block code with generator matrix~$\mathbf{G}=[\,\mathbf{I}_k \ \mathbf{P}]\in \mathbb{F}^{k\times n}$ which is $(W, B, N)$-achievable where $\mathbf{P}$ has the form~\eqref{parityLowRate}.
\\ \indent
Combining the two cases, there exists an $(n, k, T)_{\mathbb{F}}$-block code which is $(W, B, N)$-achievable.
 Based on the $(n, k, T)_{\mathbb{F}}$-block code, we can construct an $(n, k, n-1, T)_{\mathbb{F}}$-convolutional code according to Lemma~\ref{lemmaBlockToStreaming}. In addition, since $\mathbf{P}$ has the form either~\eqref{parityHighRate} or~\eqref{parityLowRate}, it follows from~\eqref{appendixAeq3} that $\mathbf{G}_\ell^{\text{conv}}=\mathbf{0}^{k\times n}$ for any $\ell\ge k+N = T+1$, which implies that the $(n, k, n-1, T)_{\mathbb{F}}$-convolutional code is also an $(n, k, T, T)_{\mathbb{F}}$-convolutional code (cf.\ Definition~\ref{definitionConvCode}).
 This concludes the proof.
\section{Proof of Lemma~\ref{lemmaHighRate}}\label{sectionLemmaHighRate}
Fix any $(W,T,B,N)$ that satisfies~\eqref{parametersFeasible} and recall that $k=T-N+1$. Suppose $k\ge B$. Fix any finite field $\mathbb{F}$ that satisfies~\eqref{assumptionFieldSize}. Our goal is to show that $\mathbf{G} = [\,\mathbf{I}_k \ \mathbf{P}]$ satisfies~\eqref{stLemmaSufficientCondition} for some $\mathbf{P}$ having the form \eqref{parityHighRate},  where $\mathbf{I}_{k-i}^{(k)}$, $\mathbf{G}_i$ and $\mathbf{E}_{\varepsilon^{T+1}}$ in~\eqref{stLemmaSufficientCondition}  are as defined in~\eqref{defMatrixI}, \eqref{defMatrixGi} and~\eqref{defMatrixE} respectively. To this end, we construct a variable vector $\vec q^{\,(i)}\triangleq [  q_0^{(i)} \  q_1^{(i)} \ \ldots \ q_{N-1}^{(i)}] \in \mathbb{F}^N$ for each $i\in\{0,1, \ldots B-1\}$ where the values of the~$B$ vectors will be determined later in this proof. In addition, we define $\vec p^{\,(i)}$ as
\begin{align}
\vec p^{\,(i)} \triangleq
\begin{cases}
[\mathbf{0}^{1\times i}\ \vec q^{\,(i)} \ \mathbf{0}^{1\times (B-N-i)}] & \text{if $0\le i\le B-N-1$,}\\
[\mathbf{0}^{1\times (B-N)}\ \vec q^{\,(i)}] & \text{if $B-N \le i \le B-1$.}
 \end{cases} \label{defMatrixPHighRate*}
 \end{align}
Construct a $(k-B)\times B$ parity matrix of a systematic MDS $(k, k-B)$-code denoted by
$
\mathbf{V}^{(k-B)\times B}
$,
which always exists because
$|\mathbb{F}| \ge 2(2T-B+1)\ge k$ by~\eqref{assumptionFieldSize}.
Then, let
 \begin{align}
 \mathbf{P}\triangleq
 \left[\begin{array}{c} \vec p^{\,(0)} \\ \vdots \\ \vec p^{\,(B-1)} \\ \hdashline\mathbf{V}^{(k-B)\times B} \end{array}\right] \label{defMatrixPHighRate}
 \end{align}
  where $\vec p^{\,(i)}$ denotes the $(i+1)^{\text{th}}$ row of $\mathbf{P}$. It can be seen that $\mathbf{P}$ has the form~\eqref{parityHighRate}. It remains to show that~$\mathbf{G}=\left[\mathbf{I}_k\ \mathbf{P}\right]$ satisfies~\eqref{stLemmaSufficientCondition} for $i=k-1, k-2, \ldots, 0$ for some $\{\vec q^{\,(i)}\}_{i=0}^{B-1}$.
\textcolor{black}{By inspecting~\eqref{stLemmaSufficientCondition} and~\eqref{defMatrixPHighRate}, we see that condition~\eqref{stLemmaSufficientCondition} depends on only $\{\vec q^{\,(B-1-j)}\}_{j=0}^{i}$ for each $i\in\{B-1, B-2, \ldots, 0\}$ and does not depend on $\{\vec q^{\,(j)}\}_{j=0}^{B-1}$ for each~$i\in\{k-1, k-2, \ldots, B\}$. In the rest of the proof, we will verify condition~\eqref{stLemmaSufficientCondition} in the order $i=k-1, k-2, \ldots, 0$, which means that we will choose $\{\vec q^{\,(i)}\}_{i=0}^{B-1}$ by choosing $\vec q^{\,(B-1)},\vec q^{\,(B-2)}, \ldots, \vec q^{\,(0)}$ sequentially.}
  Consider the following three mutually exclusive cases which will be investigated in the following three subsections respectively:
  \subsection{Case $i=k-1, k-2, \ldots, B$}\label{subsecHighRateCase1}
In this case, we have the following fact due to~\eqref{defMatrixGi} and $\mathbf{G}=\left[\mathbf{I}_k\ \mathbf{P}\right]$:
 \begin{align}
 \mathbf{G}_{B} &= \left[\begin{array}{cccc:c}\mathbf{g}_B & \mathbf{g}_{B+1}& \ldots & \mathbf{g}_{n-1} & \mathbf{0}^{k\times N}\end{array}\right] \notag\\
 &=  \left[\begin{array}{c:cc} \begin{array}{c} \mathbf{0}^{B\times (k-B)} \\ \hdashline \mathbf{I}_{k-B}
 \end{array} & \mathbf{P} & \mathbf{0}^{k\times N}\end{array}\right]. \label{defMatrixGB}
 \end{align}
  Since
 \begin{align}
  \mathbf{I}_{k-B}^{(k)}\, \mathbf{G}_{B} =\left[\begin{array}{lll} \multicolumn{3}{c}{\mathbf{0}^{B\times (T+1)}} \\ \hdashline  \mathbf{I}_{k-B} & \mathbf{V}^{(k-B)\times B} & \mathbf{0}^{(k-B)\times N} \end{array}\right] \label{eq1ProofLemmaHighRate-}
  \end{align}
by~\eqref{defMatrixGB} (recall the definition of~$\mathbf{P}$ in~\eqref{defMatrixPHighRate} and the definition of $\mathbf{I}_{k-B}^{(k)}$ in~\eqref{defMatrixI}) and any $(k-B)$ columns of $[\mathbf{I}_{k-B} \ \mathbf{V}^{(k-B)\times B}]\in \mathbb{F}^{(k-B)\times k}$ are independent due to the property of systematic MDS codes, it follows that
\begin{align}
 \textrm{space}\left(\mathbf{I}_{k-B}^{(k)}\, \mathbf{G}_{B}\, \mathbf{E}_{\varepsilon^{T+1}}\right)   
&=  \textrm{space}\left(\mathbf{I}_{k-B}^{(k)}\, \mathbf{G}_{B} \right)
 \label{eq1ProofLemmaHighRate}
 \end{align}
for any $\varepsilon^{T+1}$ with~$B$ positions of $1$'s (multiplying $E_{\varepsilon^{T+1}}$ on the right side of a matrix has the effect of zeroing $B$ columns of the multiplied matrix). Combining~\eqref{eq1ProofLemmaHighRate} and~\eqref{eq1ProofLemmaHighRate-}, we conclude that~\eqref{stLemmaSufficientCondition} holds for all $i\in\{k-1, k-2, \ldots, B\}$.
  \subsection{Case $i=B-1, B-2, \ldots, B-N$}
  \indent We will choose $\vec q^{\,(i)}$ in a recursive manner for $i=B-1, B-2, \ldots, B-N$. Suppose $i=B-j$ for some $j\in\{1, 2, \ldots, N\}$. Assume $\vec q^{\,(B-1)}, \vec q^{\,(B-2)}, \ldots, \vec q^{\,(B-j+1)}$ have been chosen such that \eqref{stLemmaSufficientCondition} holds for~$i=B-1, B-2, \ldots, B-j+1$ for any $\varepsilon^{T+1}\in\Omega_{B,N}^{T+1}$.
 Our goal is to choose $\vec q^{\,(B-j)}$ such that \eqref{stLemmaSufficientCondition} holds for~$i=B-j$. To this end, we first recognize the following fact due to~\eqref{defMatrixGi} and $\mathbf{G}=\left[\mathbf{I}_k\ \mathbf{P}\right]$:
 \begin{align}
 \mathbf{G}_{B-j} &= \left[\begin{array}{cccc:c}\mathbf{g}_{B-j} & \mathbf{g}_{B-j+1}& \ldots & \mathbf{g}_{n-1} & \mathbf{0}^{k\times (N-j)}\end{array}\right]\notag\\
 &=  \left[\begin{array}{c:cc} \begin{array}{c} \mathbf{0}^{(B-j)\times (k-B+j)}  \\ \hdashline \mathbf{I}_{k-B+j}
 \end{array} & \mathbf{P} & \mathbf{0}^{k\times (N-j)}\end{array}\right]. \label{defMatrixGB-j}
 \end{align}
Using~\eqref{defMatrixGB-j}, \eqref{defMatrixPHighRate*} and~\eqref{defMatrixPHighRate}, we obtain
 \begin{align}
 \mathbf{I}_{k-B+j}^{(k)}\, \mathbf{G}_{B-j} =
   \left[ \begin{array}{c} {\mathbf{0}^{(B-j)\times (T+1)}} \\
   \hdashline
   \begin{array}{c:c}
   \mathbf{K}^{(k-B+j)\times (k+j)} & \begin{array}{c} \mathbf{0}^{(k-B+j)\times (N-j)} \end{array}
    \end{array}
       \end{array}
   \right] \label{eq2ProofLemmaHighRate}
  \end{align}
where
\begin{align*}
\mathbf{K}^{(k-B+j)\times (k+j)}\triangleq
\left[\begin{array}{c:c}\begin{array}{c}\mathbf{I}_{k-B+j}
    \end{array}
     &
     \begin{array}{c:c} \mathbf{0}^{j\times (B-N)} & \begin{array}{c} \vec q^{\,(B-j)} \\ \vdots \\ \vec q^{\,(B-1)}  \end{array}  \\ \hdashline  \multicolumn{2}{c}{\mathbf{V}^{(k-B)\times B}}
    \end{array}
    \end{array}\right]. 
\end{align*}
By definition, we have
\begin{align}
\mathbf{K}^{(k-B+j)\times (k+j)} = \left[\begin{array}{c:c} 1 & \begin{matrix} \mathbf{0}^{1\times (k+j-N-1)} & \vec q^{\,(B-j)}\end{matrix} \\ \mathbf{0}^{(k-B+j-1)\times 1} & \mathbf{K}^{(k-B+j-1)\times (k+j-1)}\end{array}\right]. \label{defMatrixKB-j}
\end{align}
Due to the previous case in Section~\ref{subsecHighRateCase1} and the assumption in this case, the sufficient condition~\eqref{stLemmaSufficientCondition} holds for each~$i=k-1, k-2, \ldots, B-j+1$ for any $\varepsilon^{T+1}\in\Omega_{B,N}^{T+1}$, which together with~\eqref{eq2ProofLemmaHighRate} implies that
\begin{align*}
\rank\left(\left[ \begin{array}{c:c}
   \mathbf{K}^{(k-B+j-1)\times (k+j-1)} & \begin{array}{c} \mathbf{0}^{(k-B+j-1)\times (N-j+1)} \end{array}
    \end{array}
       \right]\mathbf{E}_{\varepsilon^{T+1}}\right)= k-B+j-1
\end{align*}
and hence
\begin{align}
\rank\left(\mathbf{K}^{(k-B+j-1)\times (k+j-1)}\,\mathbf{E}_{\varepsilon^{k+j-1}}\right) = k-B+j-1 \label{assumption1LemmaHighRate}
\end{align}
for any $\varepsilon^{k+j-1}\in\Omega_{B,N}^{k+j-1}$ (cf.\ Definition~\ref{definitionMaximalErasure}).
We would like to show the existence of a $\vec q^{\,(B-j)}\in\mathbb{F}^N$ such that
\begin{align}
\mathbf{u}_0^{(k-B+j)}\in \textrm{space}\Big(\mathbf{K}^{(k-B+j)\times (k+j)}\,\mathbf{E}_{\varepsilon^{k+j}}\Big)  \label{proofOfCase2*}
\end{align}
for any $\varepsilon^{k+j}\in\Omega_{B,N}^{k+j}$, which together with~\eqref{eq2ProofLemmaHighRate} will then imply that~\eqref{stLemmaSufficientCondition} holds for~$i=B-j$. Fix an arbitrary $\varepsilon^{k+j}\in\Omega_{B,N}^{k+j}$ and consider the following three subcases:\\
\textbf{Subcase $\varepsilon_0=0$:}\\ \indent
Using~\eqref{defMatrixKB-j} and the assumption that $\varepsilon_0=0$, we conclude that the first column of $\mathbf{K}^{(k-B+j)\times (k+j)}\,\mathbf{E}_{\varepsilon^{k+j}}$ is $\mathbf{u}_0^{(k-B+j)}$, which together with~\eqref{defMatrixKB-j} implies that~\eqref{proofOfCase2*} holds for any choice of $\vec q^{\,(B-j)}$. \vspace{0.04 in}\\
\textbf{Subcase $\varepsilon_0=1$ and $\sum_{\ell=0}^{k+j-1} \varepsilon_\ell=B$ with all the $1$'s in $\varepsilon^{k+j}$ occupying consecutive positions:} \vspace{0.04 in}\\ \indent
In this case, $\varepsilon^{k+j}$ equals $(\underbrace{1, \ldots, 1}_{B \text{ times}}, 0, \ldots, 0)$, and
\begin{align*}
\mathbf{K}^{(k-B+j-1)\times (k+j-1)}\mathbf{E}_{(\varepsilon_1,\ldots,\varepsilon_{k+j-1})}
\end{align*}
consists of exactly $(B-1)$ zero column vectors and $k+j-1-(B-1)=k-B+j$ non-zero column vectors, and the non-zero column vectors are denoted by $\mathbf{h}_1, \mathbf{h}_2, \ldots, \mathbf{h}_{k-B+j}$. In addition,
\begin{equation*}
\rank\big(\left[\mathbf{h}_1\ \mathbf{h}_2 \ \ldots \ \mathbf{h}_{k-B+j}\right]\big) = k-B+j-1
\end{equation*}
by~\eqref{assumption1LemmaHighRate} (due to our induction hypothesis), which implies that there exists a non-zero vector $\boldsymbol{\lambda}\triangleq [\lambda_1\ \lambda_2\ \cdots\ \lambda_{k-B+j}]^t \in \mathbb{F}^{(k-B+j)\times 1}$ such that
\begin{align}
\left[\mathbf{h}_1\ \mathbf{h}_2 \ \ldots \ \mathbf{h}_{k-B+j}\right] \boldsymbol{\lambda} = \mathbf{0}^{(k-B+j-1)\times 1}. \label{eq3ProofLemmaHighRate}
\end{align}
Since $\left[\mathbf{h}_1\ \mathbf{h}_2 \ \ldots \ \mathbf{h}_{k-B+j}\right] $ contains a $(k-B)\times (k-B+j)$ submatrix of $\mathbf{V}^{(k-B)\times B}$ (which is the parity matrix of some MDS code) where any $k-B$ columns of $\mathbf{V}^{(k-B)\times B}$ are independent, it follows that any $k-B$ columns of $\left[\mathbf{h}_1\ \mathbf{h}_2 \ \ldots \ \mathbf{h}_{k-B+j}\right] $ are independent, which implies from~\eqref{eq3ProofLemmaHighRate} that $\boldsymbol{\lambda}$ contains at least $k-B+1$ non-zero elements. Consequently, it follows from~\eqref{eq3ProofLemmaHighRate} that there exists a non-zero vector $\boldsymbol{\rho}\triangleq [\mathbf{0}^{1\times(B-1)}\ \boldsymbol{\lambda}]^t \in \mathbb{F}^{(k+j-1)\times 1}$ which contains at least $k-B+1$ non-zero elements such that
\begin{align}
\left[\begin{array}{c}  \begin{matrix} \mathbf{0}^{1\times (k+j-N-1)} & \vec q^{\,(B-j)}\end{matrix} \\ \mathbf{K}^{(k-B+j-1)\times (k+j-1)}\end{array}\right] \mathbf{E}_{(\varepsilon_1,\ldots,\varepsilon_{k+j-1})} \boldsymbol{\rho}
&= \left[\begin{array}{c}   \left[\mathbf{0}^{1\times (k+j-N-1)} \ \vec q^{\,(B-j)}\right] \mathbf{E}_{(\varepsilon_1,\ldots,\varepsilon_{k+j-1})} \boldsymbol{\rho} \\ \mathbf{0}^{(k-B+j-1)\times 1}\end{array}\right]\notag\\
&=\left[\begin{array}{c}   \left[\mathbf{0}^{1\times (k+j-N-1)} \ \vec q^{\,(B-j)}\right] \boldsymbol{\rho} \\ \mathbf{0}^{(k-B+j-1)\times 1}\end{array}\right]. \label{eq4ProofLemmaHighRate}
\end{align}
Using the fact that $\vec q^{\,(B-j)}$ is a length-$N$ variable vector and $\boldsymbol{\rho}$ contains at least $k-B+1$ non-zero elements, we claim that $ \left[\mathbf{0}^{1\times (k+j-N-1)} \ \vec q^{\,(B-j)}\right] \boldsymbol{\rho}$ is a non-zero linear function of $(q_0^{(B-j)}, q_1^{(B-j)}, \ldots, q_{N-1}^{(B-j)})$, and we let $\psi_{\varepsilon^{k+j}}^{(B-j)}(\vec q^{\,(B-j)})$ denote the non-zero linear function. To see the above claim, we can assume the contrary that $ \left[\mathbf{0}^{1\times (k+j-N-1)} \ \vec q^{\,(B-j)}\right] \boldsymbol{\rho}\equiv 0$, which implies $\boldsymbol{\lambda}$ contains at least~$N$ zeros, which together with the fact that $\boldsymbol{\lambda}$ contains at least $k-B+1$ non-zero elements leads to the conclusion that $\boldsymbol{\lambda}$ contains at least~$k-B+1+N>k-B+j$ elements, contradicting that fact that the length of $\boldsymbol{\lambda}$ is $k-B+j$. Combining~\eqref{defMatrixKB-j} and~\eqref{eq4ProofLemmaHighRate}, we conclude that~\eqref{proofOfCase2*} holds as long as $\vec q^{\,(B-j)}$ satisfies $\psi_{\varepsilon^{k+j}}^{(B-j)}(\vec q^{\,(B-j)}) \ne 0$.
\\
\textbf{Subcase $\varepsilon_0=1$ and $\sum_{\ell=0}^{k+j-1} e_\ell= N$ with no restriction on the positions of $1$'s in $\varepsilon^{k+j}$:}  \\ \indent
In this case,
\begin{align*}
\mathbf{K}^{(k-B+j-1)\times (k+j-1)}\,\mathbf{E}_{(\varepsilon_1,\ldots,\varepsilon_{k+j-1})}
\end{align*}
consists of exactly $(N-1)$ zero column vectors and $k+j-1-(N-1)=k+j-N$ non-zero column vectors, which we denote as $\mathbf{h}_1, \mathbf{h}_2, \ldots, \mathbf{h}_{k+j-N} \in \mathbb{F}^{k-B+j-1}$. Construct the following $(k-B+j)\times (k+j-N)$ submatrix of $\mathbf{K}^{(k-B+j)\times (k+j)}$:
\begin{align}
\mathbf{A}\triangleq\left[\begin{matrix} p_1& p_2 & \cdots& p_{k+j-N}\\\mathbf{h}_1 & \mathbf{h}_2 & \cdots & \mathbf{h}_{k+j-N} \end{matrix}\right] \label{defMatrixA}
\end{align}
for some $[p_1\ p_2 \ \cdots\ p_{k+j-N}]\in \mathbb{F}^{k+j-N}$ which is a subvector of $\left[\begin{matrix} \mathbf{0}^{1\times (k+j-N-1)} & \vec q^{\,(B-j)}\end{matrix}\right]$. Since the length of $[p_1\ p_2 \ \cdots\ p_{k+j-N}]$ is strictly larger than the number of zeros in $\left[\begin{matrix} \mathbf{0}^{1\times (k+j-N-1)} & \vec q^{\,(B-j)}\end{matrix}\right]$, there exists an $r\in\{1, 2, \ldots, k+j-N\}$ such that $p_r \ne 0$ where $p_r$ is an element of $\vec q^{\,(B-j)}$. Since
\begin{equation*}
\rank\big(\left[\mathbf{h}_1 \ \cdots \ \mathbf{h}_{r-1}\  \mathbf{h}_{r+1}\ \cdots \ \mathbf{h}_{k+j-N}\right]\big) = k-B+j-1
\end{equation*}
by~\eqref{assumption1LemmaHighRate} (due to the induction hypothesis), we have
\begin{align*}
 \mathbf{h}_r \in \textrm{space}\big(\left[\mathbf{h}_1 \ \cdots \ \mathbf{h}_{r-1}\  \mathbf{h}_{r+1}\ \cdots \ \mathbf{h}_{k+j-N}
\right]\big),
\end{align*}
which implies that there exist a non-zero vector $\boldsymbol{\lambda}\triangleq [\lambda_1\ \ldots \lambda_{r-1}\ 1\ \lambda_{r+1}\ \ldots\ \lambda_{k+j-N}]^t \in \mathbb{F}^{(k+j-N)\times 1}$ such that
\begin{align*}
\left[\mathbf{h}_1 \ \cdots \ \mathbf{h}_{r-1}\  \mathbf{h}_{r}\ \mathbf{h}_{r+1}\ \cdots \ \mathbf{h}_{k+j-N}\right] \boldsymbol{\lambda} = \mathbf{0}^{(k-B+j-1)\times 1},
\end{align*}
which together with~\eqref{defMatrixA} and the fact $p_r\ne 0$ implies that
\begin{align}
\mathbf{A} \boldsymbol{\lambda} = \left[\begin{matrix} \varpi_{\varepsilon^{k+j}}^{(B-j)}(\vec q^{\,(B-j)}) \\ \mathbf{0}^{(k-B+j-1)\times 1}\end{matrix}\right] \label{eq7ProofLemmaHighRate}
\end{align}
for some non-zero linear function of $(q_0^{(B-j)}, q_1^{(B-j)}, \ldots, q_{N-1}^{(B-j)})$ denoted by $\varpi_{\varepsilon^{k+j}}^{(B-j)}(\vec q^{\,(B-j)})$.
Using~\eqref{defMatrixKB-j}, \eqref{defMatrixA}, the fact that $\mathbf{A}$ consists of columns of $\mathbf{K}^{(k-B+j)\times (k+j)}$ and~\eqref{eq7ProofLemmaHighRate}, we conclude that~\eqref{proofOfCase2*} holds as long as $\vec q^{\,(B-j)}$ satisfies $\varpi_{\varepsilon^{k+j}}^{(B-j)}(\vec q^{\,(B-j)}) \ne 0$.

Combining the above three subcases, we see that for any $\varepsilon^{k+j}\in \Omega_{B,N}^{k+j}$, statement~\eqref{proofOfCase2*} holds for all $i\in\{B-1, B-2, \ldots, B-N\}$ as long as $\vec q^{\,(i)}\in\mathbb{F}^N$ satisfies $\psi_{\varepsilon^{k+j}}^{(i)}(\vec q^{\,(i)}) \ne 0$ and $\varpi_{\varepsilon^{k+j}}^{(i)}(\vec q^{\,(i)}) \ne 0$. Since the number of $\vec{\underline{q}}^{\,(i)}\in\mathbb{F}^N$ that satisfies either $\psi_{\varepsilon^{k+j}}^{(i)}(\vec{\underline{q}}^{\,(i)}) = 0$ or $\varpi_{\varepsilon^{k+j}}^{(i)}(\vec{\underline{q}}^{\,(i)}) = 0$ is less than $2|\mathbb{F}|^{N-1}$ for each $i$ and each $\varepsilon^{k+j}$ and
\begin{equation*}
\left|\Omega_{B,N}^{k+j}\right| \le {{T+1}\choose{N}} + T-B+2,
\end{equation*}
 the hypothesis~\eqref{assumptionFieldSize} guarantees the following: For each $i=B-1, B-2, \ldots, B-N$ where the vectors $\vec q^{\,(B-1)}, \ldots, \vec q^{\,(i+1)}$ have been chosen, we can always choose a $\vec q^{\,(i)}\in\mathbb{F}^N$ such that $\psi_{\varepsilon^{k+j}}^{(i)}(\vec q^{\,(i)}) \ne 0$ and $\varpi_{\varepsilon^{k+j}}^{(i)}(\vec q^{\,(i)}) \ne 0$ for all $\varepsilon^{k+j}\in \Omega_{B,N}^{k+j}$ because
 \begin{align*}
 & \frac{\left|\left\{\vec{\underline{q}}^{\,(i)}\in \mathbb{F}^N \left|\,\parbox[c]{2.2 in}{$\psi_{\varepsilon^{k+j}}^{(i)}(\vec{\underline{q}}^{\,(i)}) = 0$ or $\varpi_{\varepsilon^{k+j}}^{(i)}(\vec{\underline{q}}^{\,(i)}) = 0$ for some $\varepsilon^{k+j}\in \Omega_{B,N}^{k+j}$}\right. \right\}\right|}{\text{total number of $\vec{\underline{q}}^{\,(i)}$}} \notag\\*
 &\quad \le \frac{2\left({{T+1}\choose{N}} + T-B+2\right)|\mathbb{F}|^{N-1}}{|\mathbb{F}|^N}  \\*
 &\quad < 1.
 \end{align*}
  By induction, there exist $\vec q^{\,(B-1)}, \ldots, \vec q^{\,(B-N)}$ such that $\psi_{\varepsilon^{k+j}}^{(i)}(\vec q^{\,(i)}) \ne 0$ and $\varpi_{\varepsilon^{k+j}}^{(i)}(\vec q^{\,(i)}) \ne 0$ for all~$i\in\{B-1, \ldots, B-N\}$ and all $\varepsilon^{k+j}\in \Omega_{B,N}^{k+j}$. This together with the conclusions made in the above three subcases implies that statement~\eqref{proofOfCase2*} holds for all $i\in\{B-1, B-2, \ldots, B-N\}$ and all $\varepsilon^{k+j}\in \Omega_{B,N}^{k+j}$, which together with~\eqref{eq2ProofLemmaHighRate} implies that~\eqref{stLemmaSufficientCondition} holds for all $i\in\{B-1, B-2, \ldots, B-N\}$ and all $\varepsilon^{k+j}\in \Omega_{B,N}^{k+j}$.
\subsection{Case $i=B-N-1, B-N-2,\ldots, 0$} \label{subsectionAtTheEnd}
 Suppose $\vec q^{\,(B-1)}, \vec q^{\,(B-2)}, \ldots, \vec q^{\,(B-N)}$ have been chosen in the previous subcase. We will choose $\vec q^{\,(i)}$ in a recursive manner for $i=B-N-1, B-N-2, \ldots, 0$. Suppose $i=B-N-j$ for some $j\in\{1, 2, \ldots, B-N\}$.  Assume $\vec q^{\,(B-N-1)}, \vec q^{\,(B-N-2)}, \ldots, \vec q^{\,(B-N-j+1)}$ have been chosen such that \eqref{stLemmaSufficientCondition} holds for~$i=B-N-1, B-N-2, \ldots, B-N-j+1$ for any $\varepsilon^{T+1}\in\Omega_{B,N}^{T+1}$.
  Our goal is to choose $\vec q^{\,(B-N-j)}$ such that \eqref{stLemmaSufficientCondition} holds for~$i=B-N-j$.
%
To this end, we first use the first clause in~\eqref{defMatrixPHighRate*}, \eqref{defMatrixPHighRate} and the fact $\mathbf{G}=\left[\mathbf{I}_k\ \mathbf{P}\right]$ to obtain
 \begin{align}
  \mathbf{I}_{k-B+N+j}^{(k)}\, \mathbf{G}_{B-N-j} =
   \left[ \begin{array}{c} {\mathbf{0}^{(B-N-j)\times (T+1)}} \\
   \mathbf{J}^{(k-B+N+j)\times (T+1)}
       \end{array}
   \right] \label{eq1ProofLemmaHighRate*}
  \end{align}
where
\begin{align}
&\mathbf{J}^{(k-B+N+j)\times (T+1)}\triangleq \notag\\*
&
\left[\begin{array}{c:c}\begin{array}{c}\mathbf{I}_{k-B+N+j}
    \end{array}
     &
     \begin{array}{c} \begin{array}{lllllll}  \mathbf{0}^{1\times (B-N-j)}&  q_0^{(B-N-j)} &  q_1^{(B-N-j)} & \cdots&\cdots &  \cdots & q_{N-1}^{(B-N-j)} \\ \multicolumn{2}{r}{\mathbf{0}^{1\times (B-N-j+1)}} & q_0^{(B-N-j+1)} &  q_1^{(B-N-j+1)} & \cdots &  \cdots& q_{N-2}^{(B-N-j+1)} \\ \multicolumn{2}{r}{\ddots}&  \multicolumn{1}{r}{\ddots} &   \multicolumn{1}{r}{\ddots} &   &  \multicolumn{1}{c}{\vdots}&  \multicolumn{1}{c}{\vdots} \\
   &  \multicolumn{2}{c}{\mathbf{0}^{1\times (B-N)}}& q_0^{(B-N)} &  q_1^{(B-N)} & \cdots &  q_{N-j-1}^{(B-N)}\\
   &\multicolumn{2}{c}{\vdots}& \multicolumn{1}{c}{\vdots} &  \multicolumn{1}{c}{\vdots} & \cdots &  \multicolumn{1}{c}{\vdots}\\& \multicolumn{2}{c}{\mathbf{0}^{1\times (B-N)}}& q_0^{(B-1)} &  q_1^{(B-1)} & \cdots &  q_{N-j-1}^{(B-1)} \\ \hdashline
\multicolumn{7}{c}{\mathbf{V}^{(k-B)\times (B-j)}}  \end{array}
\end{array}
    \end{array}
\!\!\!\!\!\! \right] \label{defMatrixJ}
\end{align}
with $\mathbf{V}^{(k-B)\times (B-j)}$ being the matrix consisting of the first $B-j$ columns of $\mathbf{V}^{(k-B)\times B}$.
We would like to show the existence of a $\vec q^{\,(B-N-j)}\in\mathbb{F}^N$ such that
\begin{align}
\mathbf{u}_0^{(k-B+N+j)}\in \textrm{space}\Big(\mathbf{J}^{(k-B+N+j)\times (T+1)}\,\mathbf{E}_{\varepsilon^{T+1}}\Big)  \label{proofOfCase2**}
\end{align}
for any $\varepsilon^{T+1}\in\Omega_{B,N}^{T+1}$, which together with~\eqref{eq1ProofLemmaHighRate*} will then imply that~\eqref{stLemmaSufficientCondition} holds for~$i=B-N-j$. Fix an arbitrary $\varepsilon^{T+1}\in\Omega_{B,N}^{T+1}$ and consider the following three subcases:\\
\textbf{Subcase $\varepsilon_0=0$:}\\ \indent
Using~\eqref{defMatrixJ} and the assumption that $\varepsilon_0=0$, we conclude that the first column of $\mathbf{J}^{(k-B+N+j)\times (T+1)}\,\mathbf{E}_{\varepsilon^{T+1}}$ is $\mathbf{u}_0^{(k-B+N+j)}$, which together with~\eqref{defMatrixJ} implies that~\eqref{proofOfCase2**} holds for any choice of $\vec q^{\,(B-N-j)}$. \vspace{0.04 in}\\
\vspace{0.04 in}\\
\textbf{Subcase $\varepsilon_0=1$ and $\sum_{\ell=0}^T \varepsilon_\ell=B$ with all the $1$'s in $\varepsilon^{T+1}$ occupying consecutive positions:} \vspace{0.04 in}\\ \indent
In this case, $\varepsilon^{T+1}$ equals $(\underbrace{1, \ldots, 1}_{B \text{ times}}, 0, \ldots, 0)$ and $j\le B-N$. Using~\eqref{defMatrixJ} and the fact that $(k-B+N+j)-B \le k-B$, we see that the first $k-B+1$ non-zero columns of
$
\mathbf{J}^{(k-B+N+j)\times (T+1)}\mathbf{E}_{\varepsilon^{T+1}}
$
equals
\begin{align}
\left[\begin{array}{c}\begin{array}{c:c} \mathbf{0}^{(N+j)\times (k-B)} & \begin{matrix} q_0^{(B-N-j)}\\ \mathbf{0}^{(N+j-1)\times 1}
\end{matrix}\end{array} \\ \hdashline \mathbf{V}^* \end{array}\right] \label{defMatrixFirstK-B+1}
\end{align}
where $\mathbf{V}^*$ is a $(k-B)\times (k-B+1)$ submatrix of $[ \mathbf{I}_{k-B} \ \mathbf{V}^{(k-B)\times B}]$. Since any $(k-B)$ columns of $\mathbf{V}^*$ are independent due to the property of systematic MDS matrices, it follows from~\eqref{defMatrixFirstK-B+1} that $\left[\begin{matrix} q_0^{(B-N-j)}\\ \mathbf{0}^{(k-B+N+j-1)\times 1}
\end{matrix}\right]$ is in the column space of $
\mathbf{J}^{(k-B+N+j)\times (T+1)}\mathbf{E}_{\varepsilon^{T+1}}
$, which implies that~\eqref{proofOfCase2**} holds for any choice of $\vec q^{\,(B-N-j)}$ that satisfies $q_0^{(B-N-j)}\ne 0$. 
\\
\textbf{Subcase $\varepsilon_0=1$ and $\sum_{\ell=0}^T \varepsilon_\ell= N$ with no restriction on the positions of $1$'s in $\varepsilon^{T+1}$:}  \\ \indent
In this case,
\begin{align*}
\mathbf{J}^{(k-B+N+j)\times (T+1)}\,\mathbf{E}_{\varepsilon^{T+1}}
\end{align*}
consists of exactly~$N$ zero column vectors and $(k-B+N+j)-N=k-B+j$ non-zero column vectors. Consider
\begin{align}
\mathbf{Q}\triangleq \left[\begin{array}{c:c} \begin{matrix}
\mathbf{0}^{1\times (k-B+N+j-1)}\\ \mathbf{I}_{k-B+N+j-1}\end{matrix} & \begin{matrix} \vec q^{\,(B-N-j)}\\ \mathbf{D}
\end{matrix} \end{array}\right] \in \mathbb{F}^{(k-B+N+j)\times (k-B+2N+j-1)} \label{defMatrixQ}
\end{align}
which consists of the second to the $(k-B+N+j)^{\text{th}}$ columns and the last~$N$ columns of $\mathbf{J}^{(k-B+N+j)\times (T+1)}$ where $\mathbf{D}$ is some $(k-B+N+j-1) \times N$ matrix that is readily determined by~\eqref{defMatrixJ}. Since $\varepsilon_0=1$ and $\sum_{\ell=1}^T \varepsilon_\ell= N-1$, there exists a $(k-B+N+j) \times  (k-B+N+j)$ submatrix of~$\mathbf{Q}$ denoted by
 \begin{align}
\mathbf{B}\triangleq\left[\begin{matrix} \mathbf{h}_1 & \mathbf{h}_2 & \cdots & \mathbf{h}_{k-B+N+j} \end{matrix}\right] \label{defMatrixB}
\end{align}
such that $\mathbf{h}_1, \mathbf{h}_2, \ldots, \mathbf{h}_{k-B+N+j}$ are non-zero columns of $\mathbf{J}^{(k-B+N+j)\times (T+1)}\,\mathbf{E}_{\varepsilon^{T+1}}$. By~\eqref{defMatrixB} and the linear dependence among the column vectors of the lower $(k-B+N+j-1) \times  (k-B+N+j)$ submatrix of $\mathbf{B}$, there exist a non-zero vector $\boldsymbol{\lambda}\triangleq [\lambda_1\ \lambda_2\ \ldots\ \lambda_{k-B+N+j}]^t \in \mathbb{F}^{(k-B+N+j)\times 1}$ and a linear function of $(q_0^{(B-N-j)}, q_1^{(B-N-j)}, \ldots, q_{N-1}^{(B-N-j)})$ denoted by $\chi_{\varepsilon^{T+1}}^{(B-N-j)}(\vec q^{\,(B-N-j)})$ such that
\begin{align}
\mathbf{B} \boldsymbol{\lambda} = \left[\begin{matrix} \chi_{\varepsilon^{T+1}}^{(B-N-j)}(\vec q^{\,(B-N-j)}) \\ \mathbf{0}^{(k-B+N+j-1)\times 1}\end{matrix}\right]. \label{eq7ProofLemmaHighRate*}
\end{align}
In addition, we claim that $\chi_{\varepsilon^{T+1}}^{(B-N-j)}(\vec q^{\,(B-N-j)})$ is a non-zero function. To see this claim, we can assume the contrary that  $\chi_{\varepsilon^{T+1}}^{(B-N-j)}(\vec q^{\,(B-N-j)})\equiv 0$, which would imply that $\mathbf{B} \boldsymbol{\lambda}\equiv 0$ with $\boldsymbol{\lambda}\ne \mathbf{0}^{(k-B+N+j)\times 1}$, which together with~\eqref{defMatrixQ} and~\eqref{defMatrixB} would imply
\begin{align*}
\left[\begin{array}{c:c} \begin{matrix}
\mathbf{0}^{1\times (k-B+N+j-1)}\\ \mathbf{I}_{k-B+N+j-1}\end{matrix} & \begin{matrix} \vec q^{\,(B-N-j)}\\ \mathbf{D}
\end{matrix} \end{array}\right]\boldsymbol{\lambda^\prime}=\mathbf{0}^{(k-B+N+j)\times 1}
\end{align*}
for some $\boldsymbol{\lambda^\prime}\ne \mathbf{0}^{(k-B+2N+j-1)\times 1}$,
which together with the fact that $\vec q^{\,(B-N-j)}$ does not contain any zero would imply
the contradiction that
\begin{align}
\left[\begin{array}{c:c} \begin{matrix}
\mathbf{0}^{1\times (k-B+N+j-1)}\\ \mathbf{I}_{k-B+N+j-1}\end{matrix} & \mathbf{0}^{(k-B+N+j)\times N } \end{array}\right]\boldsymbol{\lambda^\prime}=\mathbf{0}^{(k-B+N+j)\times 1} \label{eq7ProofLemmaHighRate**}
\end{align}
for some $\boldsymbol{\lambda^\prime}\ne \mathbf{0}^{(k-B+2N+j-1)\times 1}$.
Using \eqref{defMatrixB}, the fact that $\mathbf{B}$ consists of columns of $\mathbf{J}^{(k-B+N+j)\times (T+1)}\,\mathbf{E}_{\varepsilon^{T+1}}$ and~\eqref{eq7ProofLemmaHighRate*}, we conclude that~\eqref{proofOfCase2**} holds as long as $\vec q^{\,(B-N-j)}$ satisfies $\chi_{\varepsilon^{T+1}}^{(B-j)}(\vec q^{\,(B-j)}) \ne 0$.

Combining the above three subcases, we see that for any $\varepsilon^{T+1}\in \Omega_{B,N}^{T+1}$, statement~\eqref{proofOfCase2**} holds for all $i\in\{B-N-1, B-N-2, \ldots, 0\}$ as long as $\vec q^{\,(i)}\in\mathbb{F}^N$ satisfies $q_{0}^{(i)} \ne 0$ and $\chi_{\varepsilon^{T+1}}^{(i)}(\vec q^{\,(i)}) \ne 0$. Since the number of $\vec{\underline{q}}^{\,(i)}\in\mathbb{F}^N$ that satisfies either $\underline{q}_0^{(i)} = 0$ or $\chi_{\varepsilon^{T+1}}^{(i)}(\vec{\underline{q}}^{\,(i)}) = 0$ is less than $2|\mathbb{F}|^{N-1}$ for each $i$ and each $\varepsilon^{T+1}$ and
\begin{equation*}
\left|\Omega_{B,N}^{T+1}\right| \le {{T+1}\choose{N}} + T-B+2,
\end{equation*}
 the hypothesis~\eqref{assumptionFieldSize} guarantees the following: For each $i=B-N-1, B-N-2, \ldots, 0$ where the vectors $\vec q^{\,(B-1)}, \ldots, \vec q^{\,(i+1)}$ have been chosen, we can always choose a $\vec q^{\,(i)}\in\mathbb{F}^N$ such that $q_{0}^{(i)} \ne 0$ and $\chi_{\varepsilon^{T+1}}^{(i)}(\vec q^{\,(i)}) \ne 0$ for all $\varepsilon^{T+1}\in \Omega_{B,N}^{T+1}$
  because
 \begin{align*}
 & \frac{\left|\left\{\vec{\underline{q}}^{\,(i)}\in \mathbb{F}^N \left|\,\parbox[c]{2.2 in}{$\underline{q}_0^{(i)} = 0$ or $\chi_{\varepsilon^{T+1}}^{(i)}(\vec{\underline{q}}^{\,(i)}) = 0$ for some $\varepsilon^{T+1}\in \Omega_{B,N}^{T+1}$}\right. \right\}\right|}{\text{total number of $\vec{\underline{q}}^{\,(i)}$}} \notag\\*
 &\quad \le \frac{2\left({{T+1}\choose{N}} + T-B+2\right)|\mathbb{F}|^{N-1}}{|\mathbb{F}|^N}  \\
 &\quad < 1.
 \end{align*}
 By induction, there exist $\vec q^{\,(B-N-1)}, \ldots, \vec q^{\,(0)}$ such that $q_0^{(i)} \ne 0$ and $\chi_{\varepsilon^{T+1}}^{(i)}(\vec q^{\,(i)}) \ne 0$ for all~$i\in\{B-1, \ldots, B-N\}$ and all $\varepsilon^{T+1}\in \Omega_{B,N}^{T+1}$. This together with the conclusions made in the above three subcases implies that statement~\eqref{proofOfCase2**} holds for all $i\in\{B-N-1, B-N-2, \ldots, 0\}$ and all $\varepsilon^{T+1}\in \Omega_{B,N}^{T+1}$, which together with~\eqref{eq1ProofLemmaHighRate*} implies that~\eqref{stLemmaSufficientCondition} holds for all $i\in\{B-N-1, B-N-2, \ldots, 0\}$ and all $\varepsilon^{T+1}\in \Omega_{B,N}^{T+1}$.
\subsection{Combining the three cases}
Combining the three cases studied in the preceding three subsections, we conclude that there exist $\vec q^{\,(B-1)}, \vec q^{\,(B-2)}, \ldots, \vec q^{\,(0)}$ such that~$\mathbf{G}=\left[\mathbf{I}_k\ \mathbf{P}\right]$ with $\mathbf{P}$ having the form~\eqref{defMatrixPHighRate} satisfies \eqref{stLemmaSufficientCondition} for all $i\in\{k-1, k-2, \ldots, 0\}$ and all $\varepsilon^{T+1}\in \Omega_{B,N}^{T+1}$. In particular, $\mathbf{P}$ has the form~\eqref{parityHighRate}. 


\section{Proof of Lemma~\ref{lemmaLowRate}}\label{sectionLemmaLowRate}
Fix any $(W,T,B,N)$ that satisfies~\eqref{parametersFeasible} and recall that $k=T-N+1$. Suppose $k< B$. Fix any finite field $\mathbb{F}$ that satisfies~\eqref{assumptionFieldSize}.
Our goal is to show that $\mathbf{G} = [\,\mathbf{I}_k \ \mathbf{P}]$ satisfies~\eqref{stLemmaSufficientCondition} for some $\mathbf{P}$ having the form \eqref{parityLowRate},  where $\mathbf{I}_{k-i}^{(k)}$, $\mathbf{G}_i$ and $\mathbf{E}_{\varepsilon^{T+1}}$ in~\eqref{stLemmaSufficientCondition}  are as defined in~\eqref{defMatrixI}, \eqref{defMatrixGi} and~\eqref{defMatrixE} respectively. To this end, we construct a variable vector $\vec q^{\,(i)}\triangleq [q_0^{(i)} \  q_1^{(i)} \ \ldots \ q_{N-1}^{(i)}] \in \mathbb{F}^N$ for each $i\in\{0,1, \ldots B-N-1\}$ where the values of the~$B-N$ vectors will be determined later in this proof. In addition, we let $\vec q^{\,(i)}\triangleq [\vec q_{\text{left}}^{\,(i)} \ \vec q_{\text{right}}^{\,(i)}]$ where $\vec q_{\text{left}}^{\,(i)}\triangleq[q_0^{(i)}\ q_1^{(i)}\ \ldots \ q_{B-k-1}^{(i)}]$ and $\vec q_{\text{right}}^{\,(i)}\triangleq [q_{B-k}^{(i)}\ q_{B-k+1}^{(i)}\ \ldots \ q_{N-1}^{(i)}]$, and define $\vec p^{\,(i)}\in\mathbb{F}^B$ as
\begin{align}
\vec p^{\,(i)} \triangleq
[\vec q_{\text{left}}^{\,(i)}\ \mathbf{0}^{1\times i}\ \vec q_{\text{right}}^{\,(i)} \ \mathbf{0}^{1\times (B-N-i)}]  \label{defMatrixPLowRate*}
 \end{align}
 for each $i\in\{0, 1, \ldots, B-N-1\}$.
Construct a $(k-B+N)\times N$ parity matrix of a systematic MDS $(k-B+2N, k-B+N)$-code denoted by
$\mathbf{V}^{(k-B+N)\times N}$,
which always exists because $|\mathbb{F}|\ge 2(2T-B+1)\ge k-B+2N$ by~\eqref{assumptionFieldSize}.
Let $\mathbf{V}_{\text{left}}^{(k-B+N)\times (B-k)}$ be the $(k-B+N)\times (B-k)$ matrix formed by collecting the first $B-k$ columns of $\mathbf{V}^{(k-B+N)\times N}$ and let $\mathbf{V}_{\text{right}}^{(k-B+N)\times (k-B+N)}$ be the $(k-B+N)\times (k-B+N)$ matrix formed by collecting the last $k-B+N$ columns of $\mathbf{V}^{(k-B+N)\times N}$ such that
 \begin{align*}
\left[ \mathbf{V}_{\text{left}}^{(k-B+N)\times (B-k)} \ \mathbf{V}_{\text{right}}^{(k-B+N)\times (k-B+N)}\right] =\mathbf{V}^{(k-B+N)\times N}.
 \end{align*}
Then, let
 \begin{align}
 \mathbf{P}\triangleq
 \left[\begin{array}{c} \begin{matrix} \vec p^{\,(0)} \\ \vdots \\ \vec p^{\,(B-N-1)} \end{matrix} \\ \hdashline \begin{array}{ccc} \mathbf{V}_{\text{left}}^{(k-B+N)\times (B-k)} & \mathbf{0}^{(k-B+N)\times (B-N)} &\mathbf{V}_{\text{right}}^{(k-B+N)\times (k-B+N)}\end{array}\end{array}\right]. \label{defMatrixPLowRate}
 \end{align}
  It can be seen that $\mathbf{P}$ has the form~\eqref{parityLowRate}. It remains to show that~$\mathbf{G}=\left[\mathbf{I}_k\ \mathbf{P}\right]$ satisfies \eqref{stLemmaSufficientCondition} for $i=k-1, k-2, \ldots, 0$ for some $\{\vec q^{\,(i)}\}_{i=0}^{B-N-1}$.
 \textcolor{black}{By inspecting~\eqref{stLemmaSufficientCondition} and~\eqref{defMatrixPLowRate}, we see that condition~\eqref{stLemmaSufficientCondition} depends on only $\{\vec q^{\,(B-N-1-j)}\}_{j=0}^{i}$ for each $i\in\{B-N-1, B-N-2, \ldots, 0\}$ and does not depend on $\{\vec q^{\,(i)}\}_{i=0}^{B-N-1}$ for each $i\in\{k-1, k-2, \ldots, B-N\}$. In the rest of the proof, we will verify condition~\eqref{stLemmaSufficientCondition} in the order $i=k-1, k-2, \ldots, 0$, which means that we will choose $\{\vec q^{\,(i)}\}_{i=0}^{B-N-1}$ by choosing $\vec q^{\,(B-N-1)},\vec q^{\,(B-N-2)}, \ldots, \vec q^{\,(0)}$ sequentially.}
  Consider the following two mutually exclusive cases which will be investigated in the following two subsections respectively:
  \subsection{Case $i=k-1, k-2, \ldots, B-N$} \label{subsecLowRateCase1}
In this case, we have the following fact due to~\eqref{defMatrixGi} and $\mathbf{G}=\left[\mathbf{I}_k\ \mathbf{P}\right]$:
 \begin{align}
 \mathbf{G}_{B-N} &= \left[\begin{array}{cccc}\mathbf{g}_{B-N} & \mathbf{g}_{B-N+1}& \ldots & \mathbf{g}_{n-1}\end{array}\right]\notag\\
 &=  \left[\begin{array}{c:c} \begin{array}{c} \mathbf{0}^{(B-N)\times (k-B+N)}  \\ \hdashline \mathbf{I}_{k-B+N}
 \end{array} & \mathbf{P} \end{array}\right]. \label{defMatrixGB-N}
 \end{align}
Using~\eqref{defMatrixGB-N}, the definition of~$\mathbf{P}$ in~\eqref{defMatrixPLowRate} and the definition of $ \mathbf{I}_{k-B+N}^{(k)}$ in~\eqref{defMatrixI}, we have
 \begin{align}
 \mathbf{I}_{k-B+N}^{(k)}\, \mathbf{G}_{B-N} = \left[\begin{array}{llll} \multicolumn{4}{c}{\mathbf{0}^{(B-N)\times (T+1)}} \\ \hdashline \mathbf{I}_{k-B+N} & \mathbf{V}_{\text{left}}^{(k-B+N)\times (B-k)} & \mathbf{0}^{(k-B+N)\times (B-N)} &\mathbf{V}_{\text{right}}^{(k-B+N)\times (k-B+N)}\end{array}\right]. \label{eq3ProofLemmaLowRate}
 \end{align}
Since any $(k-B+N)$ columns of
\begin{equation*}
\left[\mathbf{I}_{k-B+N} \  \mathbf{V}_{\text{left}}^{(k-B+N)\times (B-k)} \ \mathbf{V}_{\text{right}}^{(k-B+N)\times (k-B+N)}\right]\in \mathbb{F}^{(k-B+N)\times (k-B+2N)}
\end{equation*}
are independent due to the property of systematic MDS codes and
\begin{equation*}
\left[\begin{array}{lllll} \multicolumn{4}{c}{\mathbf{0}^{(B-N)\times (T+1)}} \\  \mathbf{I}_{k-B+N} & \mathbf{V}_{\text{left}}^{(k-B+N)\times (B-k)} & \mathbf{0}^{(k-B+N)\times (B-N)} &\mathbf{V}_{\text{right}}^{(k-B+N)\times (k-B+N)}\end{array}\right] \mathbf{E}_{\varepsilon^{T+1}}
\end{equation*}
 contains at least $(k-B+N)$ non-zero columns for any $\varepsilon^{T+1}\in\Omega_{B,N}^{T+1}$ with~$N$ arbitrary positions of~$1$ and exactly $(k-B+N)$ non-zero columns for any $\varepsilon^{T+1}\in\Omega_{B,N}^{T+1}$ with $B$ consecutive positions of~$1$,
it follows from~\eqref{eq3ProofLemmaLowRate} that
 \begin{align*}
 &\textrm{space}\left(\mathbf{I}_{k-B+N}^{(k)}\, \mathbf{G}_{B-N} \mathbf{E}_{\varepsilon^{T+1}}\right)\notag\\*
  &=  \textrm{space}\left(\left[\begin{array}{lllll} \multicolumn{4}{c}{\mathbf{0}^{(B-N)\times (T+1)}} \\  \mathbf{I}_{k-B+N} & \mathbf{V}_{\text{left}}^{(k-B+N)\times (B-k)} & \mathbf{0}^{(k-B+N)\times (B-N)} &\mathbf{V}_{\text{right}}^{(k-B+N)\times (k-B+N)}\end{array}\right] \mathbf{E}_{\varepsilon^{T+1}}\right)\\
 & = \textrm{space}\left(\left[\begin{array}{l} \mathbf{0}^{(B-N)\times (k-B+N)} \\ \mathbf{I}_{k-B+N} \end{array}\right] \right) 
 \end{align*}
 for any $\varepsilon^{T+1}\in\Omega_{B,N}^{T+1}$, which then implies
 \eqref{stLemmaSufficientCondition} for all $i\in\{k-1, k-2, \ldots, B-N\}$.
  \subsection{Case $i=B-N-1, B-N-2, \ldots, 0$}
  We will choose $\vec q^{\,(i)}$ in a recursive manner for $i=B-N-1, B-N-2, \ldots, 0$. Suppose $i=B-N-j$ for some $j\in\{1, 2, \ldots, B-N\}$.  Assume $\vec q^{\,(B-N-1)}, \vec q^{\,(B-N-2)}, \ldots, \vec q^{\,(B-N-j+1)}$ have been chosen such that \eqref{stLemmaSufficientCondition} holds for~$i=B-N-1, B-N-2, \ldots, B-N-j+1$ for any $\varepsilon^{T+1}\in\Omega_{B,N}^{T+1}$.
  Our goal is to choose $\vec q^{\,(B-N-j)}$ such that \eqref{stLemmaSufficientCondition} holds for~$i=B-N-j$. To this end, we first use~\eqref{defMatrixPLowRate*}, \eqref{defMatrixPLowRate} and the fact $\mathbf{G}=\left[\mathbf{I}_k\ \mathbf{P}\right]$ to obtain
 \begin{align}
  \mathbf{I}_{k-B+N+j}^{(k)}\, \mathbf{G}_{B-N-j} =
   \left[ \begin{array}{c} {\mathbf{0}^{(B-N-j)\times (T+1)}} \\
   \mathbf{J}^{(k-B+N+j)\times (T+1)}
       \end{array}
   \right] \label{eq1ProofLemmaLowRate*}
  \end{align}
where
\begin{align}
&\mathbf{J}^{(k-B+N+j)\times (T+1)}\triangleq \notag\\*
&
\left[\begin{array}{c:c:c}\begin{array}{c}\mathbf{I}_{k-B+N+j}
    \end{array}
   &
   \begin{matrix}\vec q^{\,(B-N-j)}_{\text{left}}  \\ \vdots \\ \vec q^{\,(B-N-1)}_{\text{left}} \\ \hdashline \mathbf{V}_{\text{left}}^{(k-B+N)\times (B-k)}\\
   \end{matrix}
     &
   \!\!\!\!  \begin{array}{c} \begin{array}{llllll}  \mathbf{0}^{1\times (B-N-j)}&  q_{B-k}^{(B-N-j)} & \cdots & \cdots &  \cdots & q_{N-1}^{(B-N-j)}  \\ \multicolumn{2}{c}{\ddots}&  \multicolumn{1}{r}{\ddots} &      &  \multicolumn{1}{c}{\vdots}&  \multicolumn{1}{c}{\vdots} \\
   &  \multicolumn{2}{c}{\mathbf{0}^{1\times (B-N-1)}} &  q_{B-k+1}^{(B-N-1)} & \cdots &  q_{N-j}^{(B-N-1)}\\ \hdashline
\multicolumn{4}{c}{\mathbf{0}^{(k-B+N)\times (B-N)}} &\multicolumn{2}{l}{\mathbf{V}_{\text{right}}^{(k-B+N)\times (k-B+N-j)}}  \end{array}
\end{array}
    \end{array}
\!\!\!\!\!\! \right] \label{defMatrixJLowRate}
\end{align}
with $\mathbf{V}_{\text{right}}^{(k-B+N)\times (k-B+N-j)}$ being the matrix consisting of the first $k-B+N-j$ columns of $\mathbf{V}_{\text{right}}^{(k-B+N)\times (k-B+N)}$.
We would like to show the existence of a $\vec q^{\,(B-N-j)}\in\mathbb{F}^N$ such that
\begin{align}
\mathbf{u}_0^{(k-B+N+j)}\in \textrm{space}\Big(\mathbf{J}^{(k-B+N+j)\times (T+1)}\,\mathbf{E}_{\varepsilon^{T+1}}\Big)  \label{proofOfCase2**LowRate}
\end{align}
for any $\varepsilon^{T+1}\in\Omega_{B,N}^{T+1}$, which together with~\eqref{eq1ProofLemmaLowRate*} will then imply that~\eqref{stLemmaSufficientCondition} holds for~$i=B-N-j$. Fix an arbitrary $\varepsilon^{T+1}\in\Omega_{B,N}^{T+1}$ and consider the following three subcases:\\
\textbf{Subcase $\varepsilon_0=0$:}\\ \indent
Using~\eqref{defMatrixJ} and the assumption that $\varepsilon_0=0$, we conclude that the first column of $\mathbf{J}^{(k-B+N+j)\times (T+1)}\,\mathbf{E}_{\varepsilon^{T+1}}$ is $\mathbf{u}_0^{(k-B+N+j)}$, which together with~\eqref{defMatrixJLowRate} implies that~\eqref{proofOfCase2**LowRate} holds for any choice of $\vec q^{\,(B-N-j)}$. \vspace{0.04 in}\\
\vspace{0.04 in}\\
\textbf{Subcase $\varepsilon_0=1$ and $\sum_{\ell=0}^T \varepsilon_\ell=B$ with all the $1$'s in $\varepsilon^{T+1}$ occupying consecutive positions:} \vspace{0.04 in}\\ \indent
In this case, $\varepsilon^{T+1}$ equals $(\underbrace{1, \ldots, 1}_{B \text{ times}}, 0, \ldots, 0)$ and $B\ge N+j$. Therefore, it follows from~\eqref{defMatrixJLowRate} that the first non-zero column of
$
\mathbf{J}^{(k-B+N+j)\times (T+1)}\mathbf{E}_{\varepsilon^{T+1}}
$
equals
\begin{align*}
\left[\begin{matrix} q_0^{(B-N-j)}\\ \mathbf{0}^{(k-B+N+j-1)\times 1}
\end{matrix}\right], 
\end{align*}
which implies that~\eqref{proofOfCase2**LowRate} holds for any choice of $\vec q^{\,(B-N-j)}$ that satisfies $q_0^{(B-N-j)}\ne 0$.
\\
\textbf{Subcase $\varepsilon_0=1$ and $\sum_{\ell=0}^T \varepsilon_\ell= N$ with no restriction on the positions of $1$'s in $\varepsilon^{T+1}$:}  \\ \indent
In this case,
\begin{align*}
\mathbf{J}^{(k-B+N+j)\times (T+1)}\,\mathbf{E}_{\varepsilon^{T+1}}
\end{align*}
has at least $k-B+N+j$ non-zero column vectors. Consider
\begin{align*}
\mathbf{Q}\triangleq \left[\begin{array}{c:c} \begin{matrix}
\mathbf{0}^{1\times (k-B+N+j-1)}\\ \mathbf{I}_{k-B+N+j-1}\end{matrix} & \begin{matrix} \vec q^{\,(B-N-j)}\\ \mathbf{D}
\end{matrix} \end{array}\right] \in \mathbb{F}^{(k-B+N+j)\times (k-B+2N+j-1)} 
\end{align*}
which consists of the second to the $(N+j)^{\text{th}}$ columns and the last~$N-B+k$ columns of $\mathbf{J}^{(k-B+N+j)\times (T+1)}$ where $\mathbf{D}$ is some $(k-B+N+j-1) \times N$ matrix that is readily determined by~\eqref{defMatrixJLowRate}. Since $\varepsilon_0=1$ and $\sum_{\ell=1}^T \varepsilon_\ell= N-1$, there exists a $(k-B+N+j) \times  (k-B+N+j)$ submatrix of~$\mathbf{Q}$ denoted by
 \begin{align}
\mathbf{B}\triangleq\left[\begin{matrix} \mathbf{h}_1 & \mathbf{h}_2 & \cdots & \mathbf{h}_{k-B+N+j} \end{matrix}\right] \label{defMatrixBLowRate}
\end{align}
such that $\mathbf{h}_1, \mathbf{h}_2, \ldots, \mathbf{h}_{k-B+N+j}$ are non-zero columns of $\mathbf{J}^{(k-B+N+j)\times (T+1)}\,\mathbf{E}_{\varepsilon^{T+1}}$.
By~\eqref{defMatrixBLowRate} and the linear dependence among the column vectors of the lower $(k-B+N+j-1) \times  (k-B+N+j)$ submatrix of $\mathbf{B}$, there exist a non-zero vector $\boldsymbol{\lambda}\triangleq [\lambda_1\ \lambda_2\ \ldots\ \lambda_{k-B+N+j}]^t \in \mathbb{F}^{(k-B+N+j)\times 1}$ and a linear function of $(q_0^{(B-N-j)}, q_1^{(B-N-j)}, \ldots, q_{N-1}^{(B-N-j)})$ denoted by $\chi_{\varepsilon^{T+1}}^{(B-N-j)}(\vec q^{\,(B-N-j)})$ such that
\begin{align}
\mathbf{B} \boldsymbol{\lambda} = \left[\begin{matrix} \chi_{\varepsilon^{T+1}}^{(B-N-j)}(\vec q^{\,(B-N-j)}) \\ \mathbf{0}^{(k-B+N+j-1)\times 1}\end{matrix}\right]. \label{eq7ProofLemmaLowRate*}
\end{align}
In addition, $\chi_{\varepsilon^{T+1}}^{(B-N-j)}(\vec q^{\,(B-N-j)})$ is a non-zero function by very similar arguments used in the proof of Lemma~\ref{lemmaLowRate} between \eqref{eq7ProofLemmaHighRate*} and~\eqref{eq7ProofLemmaHighRate**}.
Using \eqref{defMatrixBLowRate}, the fact that $\mathbf{B}$ consists of columns of $\mathbf{J}^{(k-B+N+j)\times (T+1)}\,\mathbf{E}_{\varepsilon^{T+1}}$ and~\eqref{eq7ProofLemmaLowRate*}, we conclude that~\eqref{proofOfCase2**LowRate} holds as long as $\vec q^{\,(B-N-j)}$ satisfies $\chi_{\varepsilon^{T+1}}^{(B-j)}(\vec q^{\,(B-j)}) \ne 0$.

Combining the above three subcases and following similar arguments used in the proof of Lemma~\ref{lemmaLowRate} at the end of Section~\ref{subsectionAtTheEnd},
we conclude that there exist $\vec q^{\,(B-N-1)}, \ldots, \vec q^{\,(0)}$ such that $q_0^{(i)} \ne 0$ and $\chi_{\varepsilon^{T+1}}^{(i)}(\vec q^{\,(i)}) \ne 0$ for all~$i\in\{B-1, \ldots, B-N\}$ and all $\varepsilon^{T+1}\in \Omega_{B,N}^{T+1}$. This together with the conclusions made in the above three subcases implies that statement~\eqref{proofOfCase2**LowRate} holds for all $i\in\{B-N-1, B-N-2, \ldots, 0\}$ and all $\varepsilon^{T+1}\in \Omega_{B,N}^{T+1}$, which together with~\eqref{eq1ProofLemmaLowRate*} implies that~\eqref{stLemmaSufficientCondition} holds for all $i\in\{B-N-1, B-N-2, \ldots, 0\}$ and all $\varepsilon^{T+1}\in \Omega_{B,N}^{T+1}$.
\subsection{Combining the two cases}
Combining the two cases studied in the preceding two subsections, we conclude that there exist $\vec q^{\,(B-N-1)}, \vec q^{\,(B-N-2)}, \ldots, \vec q^{\,(0)}$ such that~$\mathbf{G}=\left[\mathbf{I}_k\ \mathbf{P}\right]$ with $\mathbf{P}$ having the form~\eqref{defMatrixPLowRate} satisfies \eqref{stLemmaSufficientCondition} for all $i\in\{k-1, k-2, \ldots, 0\}$ and all $\varepsilon^{T+1}\in \Omega_{B,N}^{T+1}$. In particular, $\mathbf{P}$ has the form~\eqref{parityLowRate}.


\section{Optimal Convolutional Codes with Given Column Distance, Column Span and Decoding Delay}\label{sectionColumnDistanceSpan}
In this section, we will use Theorem~\ref{thmMainResult} and existing results to derive the maximum achievable rate for convolutional codes given any column distance, column span and decoding delay.
For an $(n, k, m, T)_{\mathbb{F}}$-convolutional code with memory~$m$ and generator matrices $\mathbf{G}_0^{\text{conv}}, \mathbf{G}_1^{\text{conv}}, \ldots, \mathbf{G}_{m}^{\text{conv}}$ (cf.\ Definition~\ref{definitionConvCode}),
define
\begin{align}
\mathbf{G}^{\text{conv}}\triangleq \left[\begin{array}{llll} \mathbf{G}_0^{\text{conv}}& \mathbf{G}_1^{\text{conv}}&\cdots &\mathbf{G}_T^{\text{conv}} \\ \mathbf{0}^{k\times n}& \mathbf{G}_0^{\text{conv}}& \cdots &  \mathbf{G}_{T-1}^{\text{conv}}\\\vdots & \vdots& \ddots& \vdots\\ \mathbf{0}^{k\times n}  &\mathbf{0}^{k\times n} &\ldots&  \mathbf{G}_0^{\text{conv}}\end{array}\right] \label{defGconv}
\end{align}
to be the truncated generator matrix where $\mathbf{G}_\ell^{\text{conv}}\triangleq\mathbf{0}^{k\times n}$ for any $m<\ell\le T$ by convention.
The following two definitions are standard (see, e.g., \cite[Appendix~A]{BPKTA17}).
\medskip
\begin{Definition}\label{definitionColDist}
For each $(n, k, m, T)_{\mathbb{F}}$-convolutional code, the column distance is
\begin{align*}
\mathrm{d}_T\triangleq \min\left\{ \left. \mathrm{wt}\Big([\mathbf{s}_0\ \mathbf{s}_1 \ \ldots \ \mathbf{s}_T]\,\mathbf{G}^{\text{conv}}\Big)\right| \mathbf{s}_0\ne \mathbf{0}^{1\times k}, \mathbf{s}_\ell \in\mathbb{F}^{k} \text{ for each $\ell\in\{1, 2, \ldots, T\}$}\right\}
\end{align*}
where
\begin{align*}
\mathrm{wt}\Big([\mathbf{x}_0\ \mathbf{x}_1 \ \ldots \ \mathbf{x}_{T}]\Big) \triangleq \Big|\left\{i\in\{0, 1, \ldots, T\}\left|\,\mathbf{x}_i\ne \mathbf{0}^{1\times n}\right.\right\}\Big|
\end{align*}
denotes the weight of $[\mathbf{x}_0\ \mathbf{x}_1 \ \ldots \ \mathbf{x}_{T}]$ for any $[\mathbf{x}_0\ \mathbf{x}_1 \ \ldots \ \mathbf{x}_{T}]\in \mathbb{F}^{(T+1)n}$.
\end{Definition}
\smallskip
\begin{Definition}\label{definitionColSpan}
For each $(n, k, m, T)_{\mathbb{F}}$-convolutional code, the column span is
\begin{align*}
\mathrm{c}_T\triangleq \min\left\{ \left. \textrm{space}\Big([\mathbf{s}_0\ \mathbf{s}_1 \ \ldots \ \mathbf{s}_T]\,\mathbf{G}^{\text{conv}}\Big)\right| \mathbf{s}_0\ne \mathbf{0}^{1\times k}, \mathbf{s}_\ell \in\mathbb{F}^{k} \text{ for each $\ell\in\{1, 2, \ldots, T\}$}\right\}
\end{align*}
where
\begin{align*}
&\textrm{space}\Big([\mathbf{x}_0\ \mathbf{x}_1 \ \ldots \ \mathbf{x}_{T}]\Big) \notag\\*
&\quad\triangleq
\max\left\{i\in\{0, 1, \ldots, T\}\left|\,\mathbf{x}_i\ne \mathbf{0}^{1\times n}\right.\right\} - \min\left\{i\in\{0, 1, \ldots, T\}\left|\,\mathbf{x}_i\ne \mathbf{0}^{1\times n}\right.\right\}
\end{align*}
denotes the length of the support of $[\mathbf{x}_0\ \mathbf{x}_1 \ \ldots \ \mathbf{x}_{T}]$ for any non-zero vector $[\mathbf{x}_0\ \mathbf{x}_1 \ \ldots \ \mathbf{x}_{T}]\in \mathbb{F}^{(T+1)n}$.
\end{Definition}
\smallskip

The following proposition states a well-known fact regarding the column distance and the column span for convolutional codes (see, e.g., \cite[Appendix~A]{BPKTA17}).
\begin{Proposition} \label{propositionConvDistSpan}
Any $(n, k, m, T)_{\mathbb{F}}$-convolutional code with column distance $\mathrm{d}_T$ and column span $\mathrm{c}_T$ is $(T+1, \mathrm{c}_T-1, \mathrm{d}_T-1)$-achievable. Conversely, if an $(n, k, m, T)_{\mathbb{F}}$-convolutional code is $(T+1, B, N)$-achievable, then $\mathrm{d}_T\ge N+1$ and $\mathrm{c}_T \ge B+1$.
\end{Proposition}
\smallskip

Combining Proposition~\ref{propositionConvDistSpan} and~\eqref{boundsOnCapacity}, we conclude that
\begin{align}
\frac{k}{n} \le \frac{T-\mathrm{d}_T+2}{T+\mathrm{c}_T-\mathrm{d}_T+1} \label{knownResult3}
\end{align}
for any $(n, k, m, T)_{\mathbb{F}}$-convolutional code with column distance $\mathrm{d}_T$ and column span $\mathrm{c}_T$.
Motivated by~\eqref{knownResult3}, we define the optimality of a convolutional code as follows.
\begin{Definition}\label{definitionOptimalConvCodes}
An $(n, k, m, T)_{\mathbb{F}}$-convolutional code with column distance $\mathrm{d}_T$ and column span $\mathrm{c}_T$ is said to be \emph{optimal} if
\begin{align*}
\frac{k}{n} = \frac{T-\mathrm{d}_T+2}{T+\mathrm{c}_T-\mathrm{d}_T+1}.
\end{align*}
\end{Definition}
\smallskip

Equipped with the known results for convolutional codes stated in this section, we can use Theorem~\ref{thmMainResult} to infer the following result regarding $\mathrm{d}_T$ and $\mathrm{c}_T$ for optimal convolutional codes.
\smallskip
\begin{Theorem} \label{thmColumnDistSpan}
Fix any $T$, $d$ and $c$ where $c \ge d\ge 1$, and let $\mathbb{F}$ be a finite field that satisfies
\begin{align}
\left|\mathbb{F}\right|> 2\left({{T+1}\choose{\mathrm{d}-1}} + T-c+3\right). \label{assumptionFieldSize*}
 \end{align}
  Then, there exists an optimal $(n, k, T, T)_{\mathbb{F}}$-convolutional code with column distance $\mathrm{d}_T=d$ and column span $\mathrm{c}_T=c$.
\end{Theorem}
\begin{IEEEproof}
Let $N\triangleq d-1$ and $B\triangleq c-1$. By \eqref{assumptionFieldSize*}, $\mathbb{F}$ satisfies~\eqref{assumptionFieldSize}. By Theorem~\ref{thmMainResult}, there exists an $(n, k, T, T)_{\mathbb{F}}$-convolutional code that is $(T+1,B,N)$-achievable where $k= T-N+1$ and $n= T+B-N+1$, which implies from Proposition~\ref{propositionConvDistSpan} that $\mathrm{d}_T\ge N+1$ and $\mathrm{c}_T \ge B+1$. Since
\begin{align*}
\frac{k}{n} &= \frac{T-N+1}{T+B-N+1} \\
& \ge \frac{T-\mathrm{d}_T+2}{T+B-\mathrm{d}_T+2} \\
& \ge \frac{T-\mathrm{d}_T+2}{T+\mathrm{c}_T -\mathrm{d}_T+1}
\end{align*}
by all the preceding equations in this proof, it together with~\eqref{knownResult3} implies that
\begin{align}
\frac{k}{n} = \frac{T-\mathrm{d}_T+2}{T+\mathrm{c}_T-\mathrm{d}_T+1}. \label{eq1ThmColumnDistSpan}
\end{align}
In addition, since the equations $\mathrm{d}_T\ge N+1$, $k= T-N+1$, $n=T+B-N+1$ and~\eqref{eq1ThmColumnDistSpan} imply that $\mathrm{c}_T \le B+1$, it follows that $\mathrm{c}_T= B+1$ and hence $\mathrm{d}_T= N+1$. By Definition~\ref{definitionOptimalConvCodes}, this $(n, k, T, T)_{\mathbb{F}}$-convolutional code with column distance $\mathrm{d}_T=N+1=d$ and column span $\mathrm{c}_T=B+1=c$ is optimal.
\end{IEEEproof}
\textcolor{black}{
\begin{Remark}
Regarding Theorem~\ref{thmColumnDistSpan}, if $c=d$, then the field size requirement can be relaxed to $\left|\mathbb{F}\right|\ge T+1$ due to the following. For any $\mathrm{d}_T\in\{1, 2, \ldots, n\}$, a systematic MDS $(n, k)$-code with $n\triangleq T+1$ and $k\triangleq T-\mathrm{d}_T+2$ (any $\mathrm{d}_T-1$ erasures can be recovered) always exists as long as $|\mathbb{F}|\ge n=T+1$~\cite{MacWilliamsSloane1988}, which together with Lemma~\ref{lemmaBlockToStreaming} implies the existence of an optimal $(n, k, T, T)_{\mathbb{F}}$-convolutional code such that $\mathrm{c}_T=\mathrm{d}_T$ (any $\mathrm{d}_T-1$ erasures can be recovered).
\end{Remark}}
\section{Random Code Construction} \label{sectionPracticalImplementation}
Suppose we are given a channel model which introduces packet erasures, and we would like to communicate through the channel using an optimal $(n, k, T, T)_{\mathbb{F}}$-convolutional code with column distance $\mathrm{d}_T$ and column span $\mathrm{c}_T$, where the optimality is as defined in Definition~\ref{definitionOptimalConvCodes}. If $\mathbb{F}$ satisfies
\begin{align*}
|\mathbb{F}| > 2\left({{T+1}\choose{\mathrm{d}_T-1}} + T-\mathrm{c}_T+3\right),
\end{align*}
 Theorem~\ref{thmColumnDistSpan} guarantees the existence of such an optimal convolutional code, but does not tell us how to find it efficiently. Therefore, we suggest in this section a practical method of finding optimal convolutional codes efficiently.
 To this end, we first fix any $(T,\mathrm{d}_T,\mathrm{c}_T)$ such that $T\ge \mathrm{c}_T-1\ge \mathrm{d}_T-1$, and let $W\triangleq T+1$, $B\triangleq \mathrm{c}_T-1$, $N\triangleq \mathrm{d}_T-1$, $k\triangleq T-N+1$ and $n\triangleq k+B$. In addition, we fix a finite field $\mathbb{F}$ which does not necessarily satisfy~\eqref{assumptionFieldSize}. Our goal is to find an optimal $(n, k, T, T)_{\mathbb{F}}$-convolutional code with column distance $\mathrm{d}_T$ and column span $\mathrm{c}_T$. Recall the definition of~$\mathbf{G}^{\text{conv}}$ in~\eqref{defGconv} and the definition of $\mathbf{G}_\ell^{\text{conv}}$ in Definition~\ref{definitionConvCode}. A method of constructing $\mathbf{G}^{\text{conv}}$ of an optimal convolutional code is described in the following subsection.
 \subsection{Random Encoding} \label{subsecEncodeIID}
Consider the following two steps of constructing $\{\mathbf{G}_\ell^{\text{conv}}\}_{\ell=0}^T$ and $\mathbf{G}^{\text{conv}}$ in a random manner:
 \begin{enumerate}
 \item[($\textrm{I}$)] Construct $\mathbf{G}\triangleq\left[\mathbf{I}_k\ \mathbf{P}\right]$ through randomly generating~$\mathbf{P}$ according to the following rule:
     \begin{itemize}
     \item Depending on whether $k\ge B$ or $k<B$. we generate $\mathbf{P}$ in the form either~\eqref{parityHighRate} in Lemma~\ref{lemmaHighRate} or~\eqref{parityLowRate} in Lemma~\ref{lemmaLowRate} by selecting the non-zero elements in an i.i.d.\ fashion where each element is uniformly distributed on $\mathbb{F}$.

         \end{itemize}
        Let $\mathcal{C}^{\text{block}}(\mathbf{G})$ denote the random $(n,k,T)_{\mathbb{F}}$-block code with random generator matrix $\mathbf{G}$ as constructed above.

 \item[(\textrm{II})] Based on the $(n,k,T)_{\mathbb{F}}$-block code $\mathcal{C}^{\text{block}}(\mathbf{G})$ constructed above, we construct an $(n,k,T,T)_{\mathbb{F}}$-convolutional code denoted by $\mathcal{C}_{\text{i.i.d.}}(\mathbf{G}^{\text{conv}})$ as outlined in the proof of Theorem~\ref{thmMainResult} in Section~\ref{sectionMainResultProof}, where the generator matrix $\mathbf{G}^{\text{conv}}$ is constructed according to~\eqref{appendixAeq3} in Lemma~\ref{lemmaBlockToStreaming}. If $\mathcal{C}^{\text{block}}(\mathbf{G})$ is $(W,B,N)$-achievable, it then follows from the arguments in the proof of Theorem~\ref{thmColumnDistSpan} in Section~\ref{sectionColumnDistanceSpan} that $\mathcal{C}_{\text{i.i.d.}}(\mathbf{G}^{\text{conv}})$ is an optimal $(n,k,T,T)_{\mathbb{F}}$-convolutional code with column distance $\mathrm{d}_T$ and column span $\mathrm{c}_T$.

 \end{enumerate}

\subsection{Numerical Evaluation of a Randomly Constructed Code Being Optimal}\label{subsectionProbGoodCode}
In this subsection, we would like to estimate the probability that the random code constructed according to Section~\ref{subsecEncodeIID} is optimal. To simplify notation, we let $P_{T, \mathrm{c}_T, \mathrm{d}_T}$ denote the probability of the random code $\mathcal{C}_{\text{i.i.d.}}(\mathbf{G}^{\text{conv}})$ being an optimal $(n,k,T,T)_{\mathbb{F}}$-convolutional code with column distance $\mathrm{d}_T$ and column span $\mathrm{c}_T$. Since characterizing the exact expression of $P_{T, \mathrm{c}_T, \mathrm{d}_T}$ seems intractable, we would like to estimate $P_{T, \mathrm{c}_T, \mathrm{d}_T}$ by simulation. In our simulation, $P_{T, \mathrm{c}_T, \mathrm{d}_T}$ is estimated for the following parameters of $(T, \mathrm{c}_T, \mathrm{d}_T)$: $(7, 8, 6)$, $(7, 8, 2)$, $(7, 7, 5)$, $(7, 7, 3)$, $(7, 6, 4)$, and $(7, 5, 5)$.
For each of the aforementioned parameters~$(T, \mathrm{c}_T, \mathrm{d}_T)
$, we plot the corresponding $P_{T, \mathrm{c}_T, \mathrm{d}_T}$ for $|\mathbb{F}|=3, 7, 13, 31, 61$ by generating $3000$ samples for each $|\mathbb{F}|$, and those $P_{T, \mathrm{c}_T, \mathrm{d}_T}$'s are displayed in Table~\ref{table2}. We can see from Table~\ref{table2} that the $P_{T, \mathrm{c}_T, \mathrm{d}_T}$'s are positive for a field size as small as~$7$ and they are increasing with the field size as expected.

\renewcommand{\arraystretch}{1.5}
\begin{table}[h!]
  \begin{center}
    \caption{Success probabilities $P_{T, \mathrm{c}_T, \mathrm{d}_T}$ of generating an optimal $(n,k,T,T)_{\mathbb{F}}$-convolutional code}
    \label{table2}
    \begin{tabular}{c|c|c|c|c|c|c} 
      $|\mathbb{F}|$ & $P_{7, 8, 6}$ & $P_{7, 8, 2}$ & $P_{7, 7, 5}$ & $P_{7, 7, 3}$ & $P_{7, 6, 4}$ & $P_{7, 5, 5}$\\
      \hline
      3  & 0 & 0.0617 & 0& 0.0037& 0& 0\\
      7  & 0.1290& 0.3473 & 0.0780 & 0.1390& 0.0437& 0.0060  \\
      13  &0.4643&0.5713&0.3440&0.3760&0.2263&0.1320  \\
      31  &0.7787 & 0.7910&0.6860&0.6687&0.5773&0.4980 \\
      61  &0.8950&0.8897&0.8340&0.8173&0.7667&0.7253
    \end{tabular}
  \end{center}
\end{table}


\section{Numerical Studies} \label{sectionNumerical}
The state-of-the-art MiDAS-interleaved and MiDAS-m-MDS convolutional codes have been proposed in~\cite[Sec.~IV]{BPKTA17} for the erasure channel, whose constructions involve interleaved block codes and m-MDS codes respectively. In general, convolutional codes that involve m-MDS codes require large field size that grows exponentially in~$T$ (as mentioned in~\cite[Sec.~IV-D]{BPKTA17}), hence they may not be practical for large~$T$. On the other hand, convolutional codes that are based on interleaved block codes can be implemented with practical field size. In particular, the random convolutional codes described in the previous subsection are based on interleaved block codes as illustrated in Table~\ref{table1}, which leads to low decoding complexity (comparable to decoding a block code). Since we would like to compare the performance of the random convolutional codes described in the previous subsection with existing practical convolutional codes in real-world systems, only convolutional codes based on interleaved block codes (rather than m-MDS) codes are considered in our numerical studies. More specifically, we will compare the performance of our low-complexity random codes with several practical convolutional codes including MiDAS-interleaved codes~\cite[Sec.~IV-D]{BPKTA17} and the Martinian-Sundberg code~\cite{MartinianSundberg2004} over the following two popular statistical channel models --- the GE channel~\cite{Gilbert1960,Elliott1963} and the Fritchman channel~\cite{Fritchman1967}.

\subsection{The Gilbert-Elliott channel and the Fritchman channel}
In our numerical studies, we consider the Gilbert-Elliott (GE) channel model and the Fritchman channel model as described in~\cite[Sec.~VI]{BPKTA17}. For the sake of completeness, we briefly describe the models below.

The GE channel is a two-state Markov model which consists of a good state and a bad state. In the good state, each channel packet is lost with a
probability of $\epsilon$ whereas in the bad state each channel packet
is lost with probability~$1$.
Let $\alpha$ and $\beta$ denote the transition probabilities from the
good state to the bad state and vice versa. Then, the average loss
rate of the GE channel is given by
\begin{align*}
\frac{\beta}{\alpha+\beta}\cdot \epsilon + \frac{\alpha}{\alpha+\beta}
\end{align*}
As long as the
channel stays in the bad state, the channel behaves as a burst erasure
channel. In contrast, the channel behaves like an i.i.d.\ erasure channel when the channel stays in the good state.

The Fritchman channel model consists of one good state denoted by $G$ and $M$ bad states denoted by $E_1, E_2, \ldots, E_M$. If the state equals $G$ at time~$i$, then it will transition to $E_1$ with probability $\alpha$ or stay at state~$G$ with probability $1-\alpha$ at time~$i+1$. If the state equals~$E_M$ at time~$i$, it will transition to~$G$ with probability $\beta$ or stay at state~$E_M$ with probability $1-\beta$ at time~$i+1$. If the state equals~$E_\ell$ for some $\ell\in\{1, 2, \ldots, M-1\}$, then it will transition to $E_{\ell+1}$ with probability $\beta$ or stay at state~$E_\ell$ with probability $1-\beta$ at time~$i+1$.
In the good state, each channel packet is lost with a
probability of $\epsilon$ whereas in the bad state each channel packet
is lost with probability~$1$. Fritchman and related higher-order Markov models are commonly used to model fade durations in mobile links.

\subsection{Simulation Results}
In order to compare our random code with existing codes over practical channels, we plot their loss probabilities over the GE channel and the Fritchman channel where each loss probability is generated by simulating the codes over $10^8$ channel uses. The field size is set to be $997$.

\begin{figure}[!t]
\centering
   \subfigure[Loss probability]{
        \includegraphics[width=3 in]{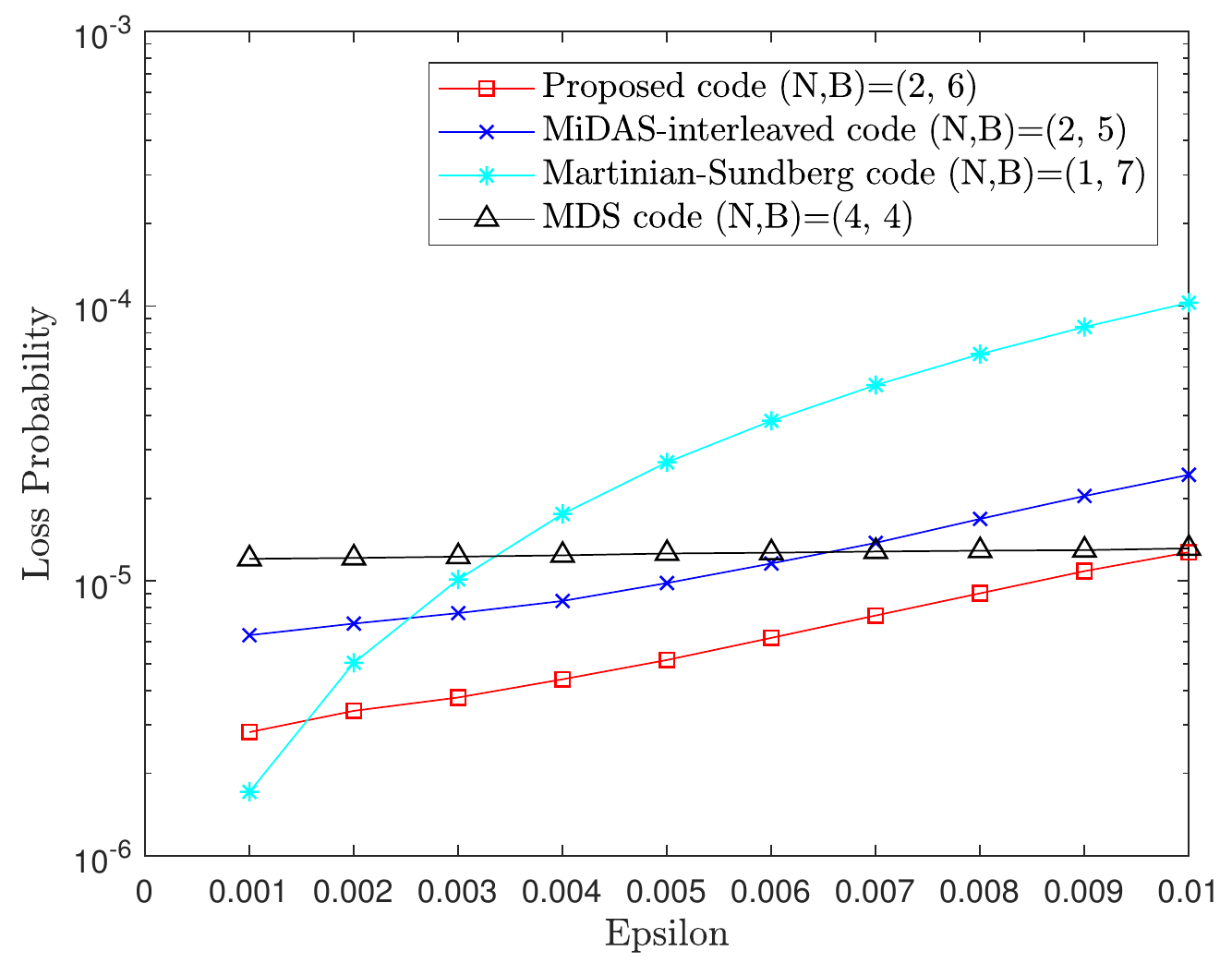}

   }
\subfigure[Burst histogram]{
        \includegraphics[width=3 in]{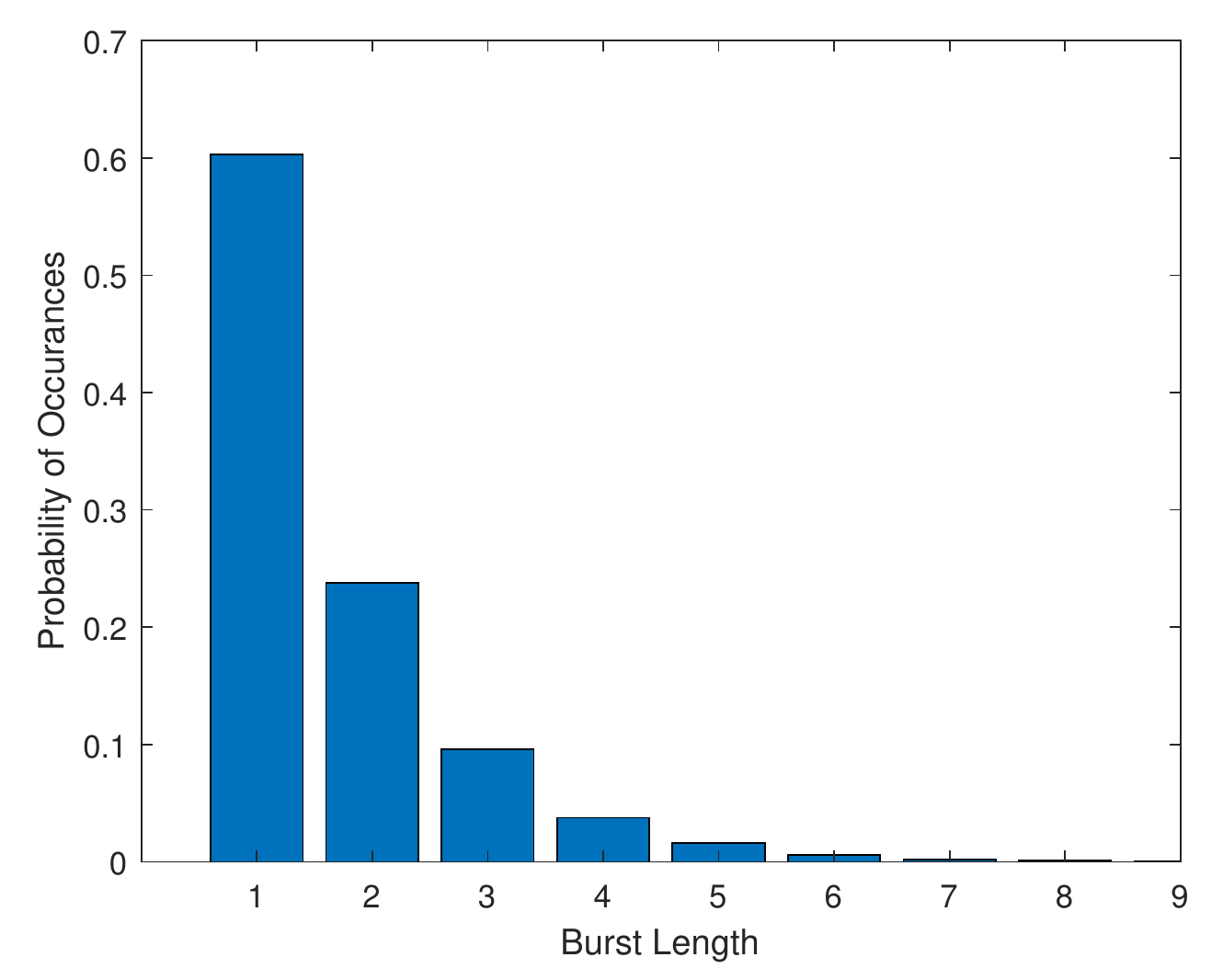}

   }
\caption{Simulation for the GE channel with $(\alpha, \beta)=(1\times 10^{-4}, 0.6)$}
\label{figure1}
\end{figure}
In Figure~\ref{figure1}(a), we plot the loss probabilities over the GE channel with constant parameters $(\alpha, \beta)=(1\times 10^{-4}, 0.6)$ and the varying parameter $\epsilon$ for our random code, the MiDAS-interleaved code, the Martinian-Sundberg code and the random MDS code with $(W,T,B,N)$ equal to
$(8, 7, 6, 2)$, $(8, 7, 5, 2)$, $(8, 7, 7, 1)$ and $(8, 7, 4, 4)$ respectively and rates equal to $1/2$, $21/41\approx 1/2$, $1/2$ and $1/2$ respectively.
The corresponding statistics of the burst length are plotted in Figure~\ref{figure1}(b), which shows that the burst histogram follows a geometric distribution with a success probability of $\beta=0.6$. As shown in Figure~\ref{figure1}(a), our random code outperforms all the other codes over the GE channel for $0.003 \le\epsilon\le 0.01$. For $\epsilon\le 0.002$, the Martinian-Sundberg code performs the best, which indicates that the loss probability in this case is dominated by burst rather than arbitrary errors. For $\epsilon>0.01$, the random MDS code performs the best, indicating that the loss probability in this case is dominated by arbitrary rather than burst errors. Indeed, our random code, the Martinian-Sundberg code and the random MDS code belong to the class of optimal convolutional codes in the sense of Definition~\ref{definitionOptimalConvCodes}. Therefore, it is not surprising that they collectively achieve the best performance as $\epsilon$ varies between~$0$ and~$1$.


\begin{figure*}[!t]
\centering
   \subfigure[Loss probability]{
        \includegraphics[width=3 in]{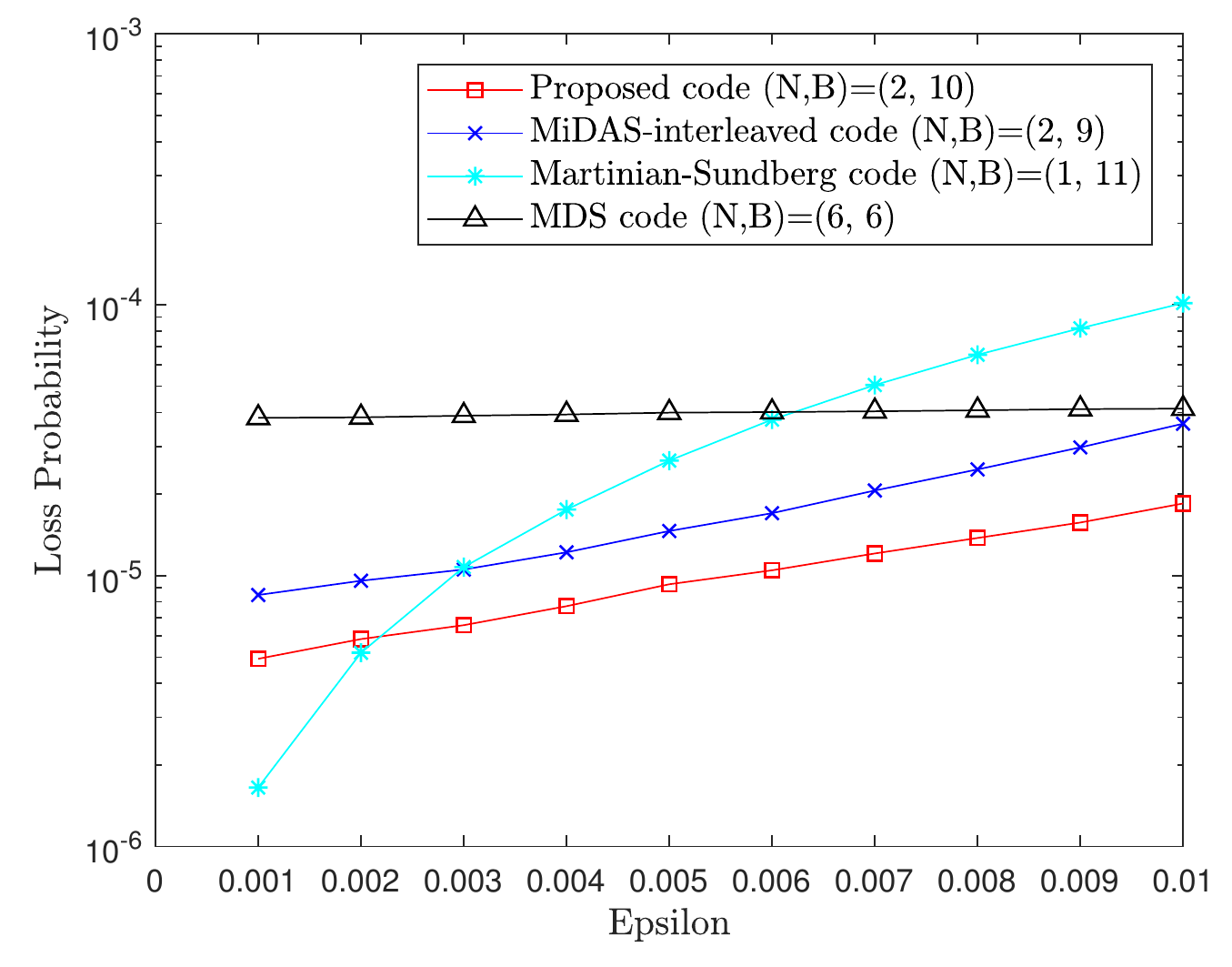}

   }
\subfigure[Burst histogram]{
        \includegraphics[width=3 in]{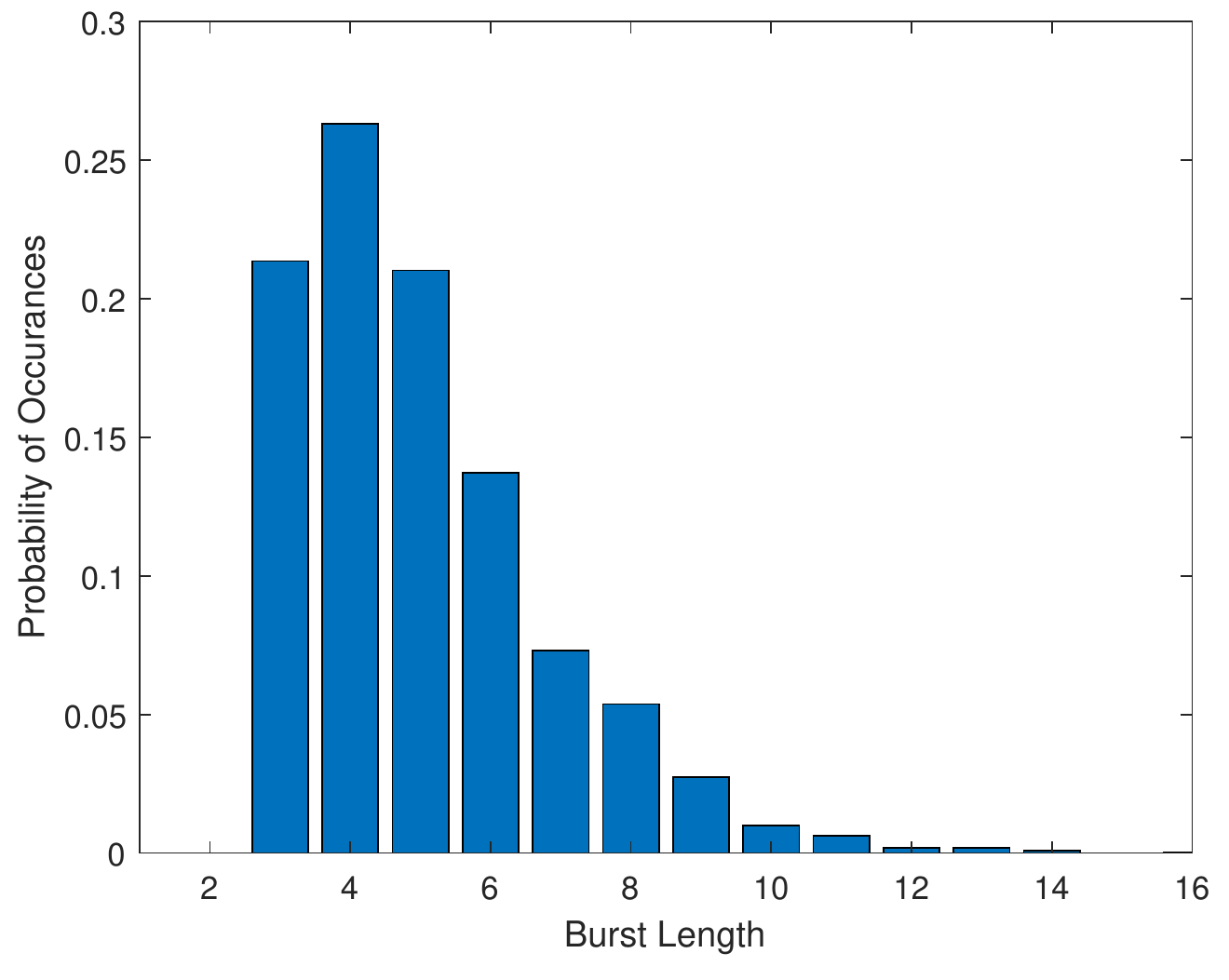}

   }
\caption{Simulation for the $4$-state Fritchman channel with $(\alpha, \beta, M)=(3\times 10^{-5}, 0.6, 4)$}
\label{figure2}
\end{figure*}
In Figure~\ref{figure2}(a), we plot the loss probabilities over the $4$-state Fritchman channel with constant parameters $(\alpha, \beta, M)=(3\times 10^{-5}, 0.6, 4)$ and the varying parameter $\epsilon$ for our random code, the MiDAS-interleaved code, the Martinian-Sundberg code and the random MDS code with $(W,T,B,N)$ equal to $(12, 11, 10, 2)$, $(12, 11, 9, 2)$, $(12, 11, 11, 1)$ and $(12, 11, 6, 6)$ respectively and rates equal to $11/21\approx 1/2$, $12/23\approx 1/2$, $12/22\approx 1/2$ and $1/2$ respectively. The corresponding statistics of the burst length are plotted in Figure~\ref{figure2}(b), which shows that the burst histogram follows a geometric distribution with a success probability of $\beta=0.5$. As shown in Figure~\ref{figure2}(a), our random code outperforms all the other codes over the $4$-state Fritchman channel for $0.003 \le\epsilon\le 0.01$. For $\epsilon\le 0.002$, the Martinian-Sundberg code performs the best, which indicates that the loss probability is dominated by burst rather than arbitrary errors. When $\epsilon$ approaches one, the random MDS code performs the best because the loss probability is dominated by arbitrary rather than burst errors. Our random code, the Martinian-Sundberg code and the random MDS code collectively achieve the best performance as $\epsilon$ varies between~$0$ and~$1$, which is consistent with the fact that they belong to the class of optimal convolutional codes in the sense of Definition~\ref{definitionOptimalConvCodes}.

\section{Concluding Remarks} \label{conclusion}
In this paper, we study streaming codes over an erasure channel whose erasure pattern in every sliding window of size~$W$ is either a burst erasure of maximum length~$B$ or multiple arbitrary erasures of maximum total count~$N$.
Under a fixed tolerable delay constraint~$T$ for each transmitted packet, we have shown in Section~\ref{subsecMainResult} the existence of convolutional codes that achieve the maximum rate of communication over the erasure channel. In addition, we have characterized in Section~\ref{sectionColumnDistanceSpan} the maximum achievable rate for convolutional codes with given column distance, column span and decoding delay. In our simulation, our proposed code outperforms all existing practical codes for various erasure probabilities over some instances of the GE channel and the Fritchman channel.

Throughout this paper, we have assumed that $W\ge T+1$ \textcolor{black}{and~\eqref{assumptionWLOG} hold} (cf.\ Section~\ref{subsecChannelModel}) and showed that the maximum achievable rate for streaming codes is $C_{(W,T,B,N)}=\frac{T-N+1}{T+B-N+1}$. For the case where $W< T+1$ \textcolor{black}{and~\eqref{assumptionWLOG} hold}, it was shown in~\cite[Th.~1]{BPKTA17} that the maximum achievable rate $C_{(W,T,B,N)}$ is bounded as
\begin{align}
 C_{(W,T,B,N)} \le \frac{W-N}{W+B-N} \label{eq1Conclusion}
\end{align}
 for any $(W, T, B, N)$. On the other hand, it follows from Theorem~\ref{thmMainResult} that
 \begin{align}
  C_{(W,W-1,B,N)}  = \frac{W-N}{W+B-N}. \label{eq2Conclusion}
 \end{align}
Since $C_{(W,T,B,N)} \ge C_{(W,W-1,B,N)} $ due to the assumption that $W<T+1$, it follows from~\eqref{eq2Conclusion} that
\begin{align}
C_{(W,T,B,N)} \ge\frac{W-N}{W+B-N}. \label{eq3Conclusion}
\end{align}
Combining~\eqref{eq1Conclusion} and~\eqref{eq3Conclusion}, we have $C_{(W,T,B,N)}=\frac{W-N}{W+B-N}$ for the case \textcolor{black}{where $W<T+1$ and~\eqref{assumptionWLOG} hold.}

\appendices
\section{Proof of Lemma~\ref{lemmaBlockToStreaming}} \label{appendixA}
Suppose we are given a $(W, B, N)$-achievable $(n, k, T)_{\mathbb{F}}$-block code, and let
$
\mathbf{G}=[\,\mathbf{I}_k \ \mathbf{P}]\in\mathbb{F}^{k\times n}
$
 be the generator matrix. By Definition~\ref{definitionAchievabilityBlock}, the $(n, k, T)_{\mathbb{F}}$-block code has the following properties:
\begin{enumerate}
\item[(i)] The length of the block code is~$n$.
\item[(ii)] From time $0$ to $k-1$, the source symbols
\begin{align*}
[x[0] \ x[1]\ \cdots \ x[k-1]]=[s[0]\ s[1]\ \cdots \ s[k-1]]
\end{align*}
     are transmitted.
    \item[(iii)] From time $k$ to $n-1$, the parity-check symbols
\begin{align*}
[x[k] \ x[k+1]\ \cdots \ x[n-1]]=[s[0]\ s[1]\ \cdots \ s[k-1]]\, \mathbf{P}
\end{align*}
     are transmitted.
 \item[(iv)] Upon receiving
 \begin{align*}
 [y[0]\ y[1]\ \ldots\ y[i+T]]=  [g_1(x[0], e_0)\ g_1(x[1], e_1)\ \ldots\ g_1(x[i+T], e_{i+T})],
 \end{align*}
  the destination can perfectly recover $s[i]$ by time $i+T$ for each $i\in\{0, 1, \ldots, k-1\}$ as long as $e^\infty\in\Omega_{(W,B,N)}^\infty$.
\end{enumerate}

In order to construct an $(n, k, n-1, T)_{\mathbb{F}}$-convolutional code, we first let $\{\mathbf{s}_i\}_{i=0}^\infty$ denote a sequence of length-$k$ packets
 and let $s_i[j]$ denote the $(j+1)^{\text{th}}$ element of $\mathbf{s}_i$ such that
\begin{align}
\mathbf{s}_i \triangleq [s_i[0]\ s_i[1]\ \cdots \ s_i[k-1] ] \label{defSiAppendixA}
\end{align}
for all $i\in\mathbb{Z}_+$. Then, construct
\begin{align}
\big[x_i[0]\ x_{i+1}[1]\ \cdots \ x_{i+n-1}[n-1] \big] \triangleq \big[s_i[0]\ s_{i+1}[1]\ \cdots \ s_{i+k-1}[k-1] \big] \mathbf{G} \label{appendixAeq1}
\end{align}
for each $i\in\mathbb{Z}_+$ where $\mathbf{G}$ is the generator matrix of the $(W, B, N)$-achievable $(n, k, T)_{\mathbb{F}}$-block code. In other words, we are coding $\mathbf{s}_i$ diagonally as illustrated in Table~\ref{table1}.
At each time $i\in\mathbb{Z}_+$, the source transmits
\begin{align}
\mathbf{x}_i\triangleq \big[x_i[0]\ x_i[1]\ \cdots \ x_i[n-1]\big]. \label{defXiAppendixA}
\end{align}
In order to express $\mathbf{x}_i$ in the form of~\eqref{xiConvForm}, we let $g_{i,j}$ be the entry situated in row~$i$ and column~$j$ of $\mathbf{G}$ such that
\begin{align*}
\mathbf{G}= \big[g_{i,j}\big]_{\substack{0\le i\le k-1,\\ 0\le j\le n-1}} %
\end{align*}
and define $\mathbf{G}_\ell^{\text{conv}}$ as in~\eqref{appendixAeq3}
for each $\ell\in\{0, 1, \ldots, n-1\}$ such that
\begin{align*}
\mathbf{G} = [\,\mathbf{I}_k \ \mathbf{P}]= \sum_{\ell=0}^{n-1} \mathbf{G}_\ell^{\text{conv}}.  
\end{align*}
Following~\eqref{defXiAppendixA}, we consider
\begin{align}
\mathbf{x}_i &= \sum_{\ell=0}^{n-1} \left[\begin{matrix}s_{i-\ell}[0]&s_{i-1-\ell}[1]&\cdots & s_{i-k+1-\ell}[k-1]\end{matrix}\right]\, \left[\begin{array}{c:c:c} \mathbf{0}^{k\times \ell} & \begin{matrix}g_{0,\ell}\\ \vdots \\ g_{k-1,\ell}\end{matrix}& \mathbf{0}^{k\times (n-\ell-1)} \end{array}\right] \label{appendixAeq4a}\\
& = \sum_{\ell=0}^{n-1} \sum_{j=0}^{k-1}\left[\begin{matrix}  \mathbf{0}^{1\times \ell} & s_{i-\ell}[j]\,g_{j,\ell} &\mathbf{0}^{1\times (n-\ell-1)}\end{matrix}\right] \notag\\
& = \sum_{\ell=0}^{n-1} \mathbf{s}_{i-\ell}\, \mathbf{G}_\ell^{\text{conv}} \label{appendixAeq4}
\end{align}
for each $i\in\mathbb{Z}_+$, where
\begin{itemize}
\item \eqref{appendixAeq4a} is due to~\eqref{appendixAeq1} and~\eqref{defXiAppendixA}.
\item \eqref{appendixAeq4} is due to~\eqref{defSiAppendixA}, \eqref{appendixAeq3}, the fact that $\mathbf{G}$ has the form $[\,\mathbf{I}_k \ \mathbf{P}]$, and the convention that $\mathbf{s}_t=\mathbf{0}^{1\times k}$ for all $t<0$.
\end{itemize}
 Based on the $(W, B, N)$-achievable $(n, k, T)_{\mathbb{F}}$-block code which satisfies Properties (i) to (iv) as stated at the beginning of this proof, we construct an $(n, k, n-1, T)_{\mathbb{F}}$-convolutional code whose encoding function at time~$i$ is specified by~\eqref{appendixAeq4}, where $\mathbf{x}_i$ and $\mathbf{s}_i$ satisfy~\eqref{defXiAppendixA} and~\eqref{defSiAppendixA} respectively. Our goal is to show that the convolutional code is $(W, B, N)$-achievable. To this end, we fix any $i\in\mathbb{Z}_+$ and any $e^\infty\in\Omega_{(W,B,N)}^\infty$, and would like to show that the destination can perfectly recover $\mathbf{s}_i=\big[s_i[0]\ s_i[1]\ \cdots \ s_i[k-1]\big]$ based on
  \begin{align}
 [\mathbf{y}_0\ \mathbf{y}_1\ \ldots\ \mathbf{y}_{i+T}]=  [g_n(\mathbf{x}_0, e_0)\ g_n(\mathbf{x}_1, e_1)\ \ldots\ g_n(\mathbf{x}_{i+T}, e_{i+T})]. \label{appendixAeq5}
 \end{align}
According to~\eqref{defXiAppendixA}, for each $i\in\mathbb{Z}_+$, $\big[x_i[0]\ x_{i+1}[1]\ \cdots \ x_{i+n-1}[n-1]\big]$ are transmitted from time~$i$ to time $i+n-1$. Therefore, it follows from~\eqref{appendixAeq1}, Property~(iv) and~\eqref{appendixAeq5} that for each $i\in\mathbb{Z}_+$ and each $\ell\in\{0, 1, \ldots, k-1\}$, the destination can perfectly recover $s_i[\ell]$ by time $i+T$ based on $ [\mathbf{y}_i\ \mathbf{y}_{i+1}\ \ldots\ \mathbf{y}_{i+T}]$, which implies that the destination can perfectly recover $\mathbf{s}_i$ time $i+T$ based on $ [\mathbf{y}_0\ \mathbf{y}_1\ \ldots\ \mathbf{y}_{i+T}]$. Consequently, for any $i\in\mathbb{Z}_+$ and any $e^\infty\in\Omega_{(W,B,N)}^\infty$, the destination can perfectly recover $\mathbf{s}_i$ by time $i+T$, which implies that the $(n, k, n-1, T)_{\mathbb{F}}$-convolutional code is $(W, B, N)$-achievable.
In addition, using~\eqref{defSiAppendixA}, \eqref{appendixAeq1} and~\eqref{defXiAppendixA}, we obtain~\eqref{st2LemmaBlockToStreaming}.


\section{Proof of Lemma~\ref{lemmaSufficientCondition}} \label{appendixB}
Let~$\mathbf{G}=[\mathbf{g}_0\ \mathbf{g}_1\ \ldots \ \mathbf{g}_{n-1}]\in\mathbb{F}^{k\times n}$ be a matrix that satisfies~\eqref{stLemmaSufficientCondition} for each~$i\in\{0, 1, \ldots, k-1\}$ and each maximal $(T+1,B,N)$-erasure pattern $\varepsilon^{T+1}\in \Omega_{B,N}^{T+1}$. First, we would like to construct an $(n, k, T)_{\mathbb{F}}$-block code with generator matrix~$\mathbf{G}$ that is $(T+1, B, N)$-achievable (cf.\ Definition~\ref{definitionAchievabilityBlock}), and this lemma will then follow because any $(T+1, B, N)$-achievable $(n, k, T)_{\mathbb{F}}$-block code is also an $(W, B, N)$-achievable $(n, k, T)_{\mathbb{F}}$-block code due to the assumption that $W\ge T+1$. Since the encoding strategy of an $(n, k, T)_{\mathbb{F}}$-block code with generator matrix~$\mathbf{G}$ is completely determined by~\eqref{encodingOfBlockCode}, it suffices to show the existence of $\{\varphi_{i+T}\}_{i=0}^{k-1}$ such that
\textcolor{black}{ \begin{align}
s[i] =
\begin{cases}
\varphi_{i+T}(g_1(x[0], e_0), g_1(x[1], e_1), \ldots, g_1(x[i+T], e_{i+T})) & \text{if $0\le i\le B-N$,}   \\           
        \varphi_{i+T}(g_1(x[0], e_0), g_1(x[1], e_1), \ldots, g_1(x[n], e_{n}))&  \text{if $B-N+1\le i\le k-1$}
\end{cases}\label{eq1AppendixA}
 \end{align}
 }
holds for any $(T+1, B, N)$-erasure sequence~$e^\infty\in \Omega_{(T+1,B,N)}^\infty$. Recognizing the fact due to~\eqref{encodingOfBlockCode} and~\eqref{defChannelLawBlock} that
\begin{align}
[g_1(x[0], e_0)\ g_1(x[1], e_1)\ \ldots\ g_1(x[n-1], e_{n-1})] \,\mathbf{E}_{e^{n}} &=[x[0]\ x[1]\ \ldots\ x[n-1]] \,\mathbf{E}_{e^{n}} \notag \\
& = [s[0]\ s[1]\ \ldots\ s[k-1]]\, \mathbf{G} \,\mathbf{E}_{e^{n}}, \label{eq2AppendixA}
\end{align}
we fix an arbitrary $(T+1, B, N)$-erasure sequence~$e^\infty\in \Omega_{(T+1,B,N)}^\infty$ and would like to show that
\begin{align}
\mathbf{u}_i^{(k)} \in
\begin{cases}
\textrm{space}\Big(\left[\mathbf{g}_0\ \mathbf{g}_1\ \cdots\ \mathbf{g}_{i+T}\right] \,\mathbf{E}_{e^{i+T+1}} \Big) & \text{if $0\le i\le B-N$,}\\
\textrm{space}\Big(\left[\mathbf{g}_0\ \mathbf{g}_1\ \cdots\ \mathbf{g}_{n-1}\right] \,\mathbf{E}_{e^{n}} \Big) & \text{if $B-N+1\le i\le k-1$}
\end{cases} \label{inductionStatementAppendixA}
\end{align}
for each $i$, which together with~\eqref{eq2AppendixA} would then imply the existence of~$\{\varphi_{i+T}\}_{i=0}^{k-1}$ that satisfy~\eqref{eq1AppendixA} for each $i$.
We will show~\eqref{inductionStatementAppendixA} by induction on $i=0, 1, \ldots, k-1$. For $i=0$, \eqref{inductionStatementAppendixA} follows directly from~\eqref{stLemmaSufficientCondition} by setting $\varepsilon^{T+1}= e^{T+1}$. Suppose~\eqref{inductionStatementAppendixA} holds for each $i=0, 1, \ldots, j$ for some $j<k-1$. Then, 
showing~\eqref{inductionStatementAppendixA} for $i^*=j+1$ is equivalent to showing
\begin{align}
\mathbf{u}_{i^*}^{(k)} \in
\begin{cases}
\textrm{space}\left(\mathbf{I}_{k-i^*}^{(k)}\,\left[\mathbf{g}_0\ \mathbf{g}_1\ \cdots\ \mathbf{g}_{i^*+T}\right] \,\mathbf{E}_{e^{i^*+T+1}} \right) & \text{if $0\le i^*\le B-N$,}\\
\textrm{space}\left(\mathbf{I}_{k-i^*}^{(k)}\,\left[\mathbf{g}_0\ \mathbf{g}_1\ \cdots\ \mathbf{g}_{n-1}\right] \,\mathbf{E}_{e^{n}} \right) & \text{if $B-N+1\le i^*\le k-1$,}
\end{cases} \label{eq3AppendixA}
\end{align}
which is a direct consequence of~\eqref{stLemmaSufficientCondition}.
By mathematical induction, we have proved that~\eqref{inductionStatementAppendixA} holds for each $i=0, 1, \ldots, k-1$.

\section*{Acknowledgment}
The authors would like to thank the Associate Editor Prof. Anxiao Jiang and the three anonymous reviewers for the useful comments that help us improve the presentation of this paper.


\begin{thebibliography}{10}
\providecommand{\url}[1]{#1}
\csname url@samestyle\endcsname
\providecommand{\newblock}{\relax}
\providecommand{\bibinfo}[2]{#2}
\providecommand{\BIBentrySTDinterwordspacing}{\spaceskip=0pt\relax}
\providecommand{\BIBentryALTinterwordstretchfactor}{4}
\providecommand{\BIBentryALTinterwordspacing}{\spaceskip=\fontdimen2\font plus
\BIBentryALTinterwordstretchfactor\fontdimen3\font minus
  \fontdimen4\font\relax}
\providecommand{\BIBforeignlanguage}[2]{{%
\expandafter\ifx\csname l@#1\endcsname\relax
\typeout{** WARNING: IEEEtran.bst: No hyphenation pattern has been}%
\typeout{** loaded for the language `#1'. Using the pattern for}%
\typeout{** the default language instead.}%
\else
\language=\csname l@#1\endcsname
\fi
#2}}
\providecommand{\BIBdecl}{\relax}
\BIBdecl

\bibitem{5GPPP}
\BIBentryALTinterwordspacing
5G-PPP, ``{5G} empowering vertical industries,'' Tech. Rep., Feb. 2015.
  [Online]. Available: \url{https://5g-ppp.eu/roadmaps/}
\BIBentrySTDinterwordspacing

\bibitem{onewayTransTime}
{International Telecommunication Union}, ``Recommendation {G.114},'' Tech.
  Rep., May 2003.

\bibitem{StockhammerHannuksela2005}
T.~Stockhammer and M.~Hannuksela, ``{H.264/AVC} video for wireless
  transmission,'' \emph{{IEEE} Wireless Commun. Mag.}, vol.~12, pp. 6--13, Aug.
  2005.

\bibitem{BKTAmagazine17}
A.~K. A.~Badr, W.-T. Tan, and J.~Apostolopoulos, ``Perfecting protection for
  interactive multimedia: {A} survey of forward error correction for low-delay
  interactive applications,'' \emph{IEEE Signal Processing Magazine}, vol.~34,
  pp. 95 -- 113, 2017.

\bibitem{HasslingerHohlfeld2008}
G.~Hasslinger and O.~Hohlfeld, ``The {Gilbert-Elliott} model for packet loss in
  real time services on the {Internet},'' in \emph{Proc.\ 14th GI/ITG
  Conference -- Measuring, Modelling and Evaluation of Computer and
  Communication Systems (MMB)}, Dortmund, Germany, Mar./Apr. 2008.

\bibitem{HGH2008}
O.~Hohlfeld, R.~Geib, and G.~Hasslinger, ``Packet loss in real-time services:
  {Markovian} models generating {QoE} impairments,'' in \emph{Proc.\ 16th
  Internatioal Workshop on Quality of Service}, Enschede, Netherlands, Jun.
  2008.

\bibitem{Gilbert1960}
E.~N. Gilbert, ``Capacity of a burst-noise channel,'' \emph{Bell System
  Technical Journal}, vol.~39, pp. 1253--–1265, Sep. 1960.

\bibitem{Elliott1963}
E.~O. Elliott, ``Estimates of error rates for codes on burst-noise channels,''
  \emph{Bell System Technical Journal}, vol.~42, pp. 1977--1997, Sep. 1963.

\bibitem{Fritchman1967}
B.~D. Fritchman, ``A binary channel characterization using partitioned {Markov}
  chains,'' \emph{{IEEE} Trans. Inf. Theory}, vol.~13, no.~2, pp. 221--227,
  1967.

\bibitem{BPKTA17}
A.~Badr, P.~Patil, A.~Khisti, W.-T. Tan, and J.~Apostolopoulos, ``Layered
  constructions for low-delay streaming codes,'' \emph{{IEEE} Trans. Inf.
  Theory}, vol.~63, no.~1, pp. 111 -- 141, 2017.

\bibitem{Hagelbarger1959}
D.~W. Hagelbarger, ``Recurrent codes: {Easily} mechanized, burst-correcting,
  binary codes,'' \emph{Bell Sys. Tech. J.}, vol.~38, pp. 969 -- 984, 1959.

\bibitem{WynerAsh1963}
A.~D. Wyner and R.~Ash, ``Analysis of recurrent codes,'' \emph{{IEEE} Trans.
  Inf. Theory}, vol.~9, pp. 143 -- 156, 1963.

\bibitem{Massey1965}
J.~L. Massey, ``Implementation of burst-correcting convolutional codes,''
  \emph{{IEEE} Trans. Inf. Theory}, vol.~11, no.~3, pp. 416 -- 422, 1965.

\bibitem{Forney1971}
G.~D. Forney, ``Burst-correcting codes for the classic bursty channel,''
  \emph{{IEEE} Trans. Inf. Theory}, vol.~19, no.~5, pp. 772 -- 781, 1971.

\bibitem{MartinianSundberg2004}
E.~Martinian and C.-E.~W. Sundberg, ``Burst erasure correction codes with low
  decoding delay,'' \emph{{IEEE} Trans. Inf. Theory}, vol.~50, no.~10, pp. 2494
  -- 2502, 2004.

\bibitem{LeungHo2012}
D.~Leong and T.~Ho, ``Erasure coding for real-time streaming,'' in \emph{Proc.
  IEEE Intl. Symp. Inf.~Theory}, Cambridge, MA, Jul. 2012.

\bibitem{LQH2013}
D.~Leong, A.~Qureshi, and T.~Ho, ``On coding for real-time streaming under
  packet erasures,'' in \emph{Proc. IEEE Intl. Symp. Inf.~Theory}, Istanbul,
  Turkey, Jul. 2013.

\bibitem{BKTA2013}
A.~Badr, A.~Khisti, W.-T. Tan, and J.~Apostolopoulos, ``Streaming codes for
  channels with burst and isolated erasures,'' in \emph{Proc.\ IEEE INFOCOM},
  Turin, Italy, Apr. 2013.

\bibitem{AdlerCassuto2017}
N.~Adler and Y.~Cassuto, ``Burst-erasure correcting codes with optimal average
  delay,'' \emph{{IEEE} Trans. Inf. Theory}, vol.~63, no.~5, pp. 2848--2865,
  2017.

\bibitem{MacWilliamsSloane1988}
F.~J. MacWilliams and N.~J.~A. Sloane, \emph{The Theory of Error-Correcting
  Codes}, 1st~ed.\hskip 1em plus 0.5em minus 0.4em\relax Amsterdam, Holland:
  North-Holland, Netherlands, 1988.

\bibitem{BKTZA2017}
A.~Badr, A.~Khisti, W.-T. Tan, X.~Zhu, and J.~Apostolopoulos, ``{FEC} for
  {VoIP} using dual-delay streaming codes,'' in \emph{Proc.\ IEEE INFOCOM},
  Atlanta, GA, USA, May 2017.

\bibitem{JohannessonZigangirovBook}
R.~Johannesson and K.~S. Zigangirov, \emph{Fundamentals of Convolutional
  Coding}, 2nd~ed.\hskip 1em plus 0.5em minus 0.4em\relax Hoboken, NY: John
  Wiley and Sons, Inc., 2015.

\end{thebibliography}


\end{document}